\documentclass[conference]{IEEEtran}

\usepackage{cite}

\ifCLASSINFOpdf
  \usepackage[pdftex]{graphicx}
  \graphicspath{{./figs/}}
  \DeclareGraphicsExtensions{.pdf,.eps}
\else
\fi

\usepackage{amsmath}
\usepackage{amsfonts}
\usepackage{mathtools}
\usepackage{cases}
\usepackage{physics}
\usepackage{booktabs}

\usepackage{algorithmic}
\renewcommand{\algorithmiccomment}[1]{\bgroup\hfill\{~#1~\}\egroup}

\usepackage{array}

\ifCLASSOPTIONcompsoc
  \usepackage[caption=false,font=normalsize,labelfont=sf,textfont=sf]{subfig}
\else
  \usepackage[caption=false,font=footnotesize]{subfig}
\fi

\usepackage[utf8]{inputenc}
\usepackage[T1]{fontenc}
\usepackage[usenames,dvipsnames,table]{xcolor}
\usepackage{diagbox}
\usepackage{multirow}
\usepackage{hyperref}

\newcommand{\ud}{\,\mathrm{d}}

\begin{document}

\title{Computing and Compressing\\
Electron Repulsion Integrals on FPGAs}

\author{%
  \IEEEauthorblockN{%
    Xin Wu\IEEEauthorrefmark{1}\IEEEauthorrefmark{2},
    Tobias Kenter\IEEEauthorrefmark{1}\IEEEauthorrefmark{2},
    Robert Schade\IEEEauthorrefmark{1}\IEEEauthorrefmark{2},
    Thomas D. Kühne\IEEEauthorrefmark{3},
    Christian Plessl\IEEEauthorrefmark{1}\IEEEauthorrefmark{2}
  }
  \IEEEauthorblockA{%
    \IEEEauthorrefmark{1}Paderborn Center for Parallel Computing, %
    \IEEEauthorrefmark{2}Department of Computer Science, %
    \IEEEauthorrefmark{3}Department of Chemistry, \\
      Paderborn University, Warburger Str. 100, 33098 Paderborn, Germany\\
      Email: \{xin.wu, tobias.kenter, robert.schade, thomas.kuehne, christian.plessl\}@uni-paderborn.de
  }
}

\maketitle

\thispagestyle{plain}
\pagestyle{plain}

\begin{abstract}
The computation of electron repulsion integrals (ERIs) over Gaussian-type orbitals (GTOs) is a challenging problem in quantum-mechanics-based atomistic simulations.
In practical simulations, several trillions of ERIs may have to be computed for every time step.

In this work, we investigate FPGAs as accelerators for the ERI computation.
We use template parameters, here within the Intel oneAPI tool flow, to create customized designs for 256 different ERI quartet classes, based on their orbitals.
To maximize data re-use, all intermediates are buffered in FPGA on-chip memory with customized layout. The pre-calculation of intermediates also helps to overcome data dependencies caused by multi-dimensional recurrence relations.
The involved loop structures are partially or even fully unrolled for high throughput of FPGA kernels.
Furthermore, a lossy compression algorithm utilizing arbitrary bitwidth integers is integrated in the FPGA kernels. %
To our best knowledge, this is the first work on ERI computation on FPGAs that supports more than just the single most basic quartet class. Also, the integration of ERI computation and compression it a novelty that is not even covered by CPU or GPU libraries so far.

Our evaluation shows that using 16-bit integer for the ERI compression, the fastest FPGA kernels exceed the performance of 10 GERIS ($10\times10^{9}$ ERIs per second) on one Intel Stratix 10 GX 2800 FPGA, with maximum absolute errors around $10^{-7}$ - $10^{-5}$ Hartree.
The measured throughput can be accurately explained by a performance model.
The FPGA kernels deployed on 2 FPGAs outperform similar computations using the widely used libint reference on a two-socket server with 40 Xeon Gold 6148 CPU cores of the same process technology by factors up to 6.0x and on a new two-socket server with 128 EPYC 7713 CPU cores by up to 1.9x.
\end{abstract}

\IEEEpeerreviewmaketitle

\section{Introduction}\label{sec:intro}

Quantum-mechanics-based atomistic simulations, known as \textit{ab initio} molecular dynamics (AIMD)~\cite{Kuehne2014}, solve the electronic structure problem for solids and molecular systems~\cite{pople98}, e.g.\ with density functional theory (DFT)~\cite{kohn98}.
AIMD can describe complex reactive systems, which are normally inaccessible by means of classical molecular dynamics.
The current state-of-the-art AIMD simulations have been pushed beyond the boundaries of more than 100 million atoms by using the novel non-orthogonalized local submatrix method~\cite{Lass20, Schade22}.
The hybrid DFT~\cite{Becke93}, which incorporates a portion of nonlocal Hartree-Fock exchange contributions, is essential for the accurate simulation of challenging systems. An example is the description of the band gap in bulk silicon used in the semiconductor industry. The band gap describes the electrical conductivity properties and is underestimated in semi-local DFT calculations. The use of hybrid DFT is required to properly obtain the band gap in agreement with experimental values~\cite{PhysRevApplied.8.024023} and, thus, properly describe the electrical conductivity.
However, AIMD simulations based on the hybrid DFT are usually limited to systems of thousands of atoms because of the computational challenges of electron repulsion integrals (ERIs) over Gaussian-type orbitals (GTOs).
The formal complexity of the ERI computation is $\mathcal{O}(n^{4})$, where $n$ is the number of atoms in the system of interest.
In practical terms, many trillions of ERIs may have to be computed in each time step of an AIMD simulation~\cite{Kuehne2020}.
Thus the ERI computation was dubbed as ``the nightmare of the integrals'' in the 1998 Nobel Lecture for Chemistry~\cite{pople98}.

Because of the importance of ERI computation for AIMD simulations, there are active endeavors to develop fast and efficient methods~\cite{HJO_ch9}, as well as libraries on CPU~\cite{libint2, libcint, Chow16, Chow18, Neese22} and GPU~\cite{Martínez08, Gordon10, Merz13, Gordon22} in last several decades.
However, the computations of necessary intermediates involve data dependencies due to multi-dimensional recurrence relations (RRs).
Furthermore, a large number of combinations of angular momenta for both recursive intermediates and final ERIs require flexible data layouts.
The hardware architectures of CPU and GPU are all fixed at the time of fabrication, thus cannot fulfill such flexible requirements for the ERI computation owing to either unfilled SIMD vectors on CPU~\cite{Chow16}, or restrictions on the number of GPU threads per block and limited shared memory per streaming processor~\cite{Gordon22}.
In contrast to thread-level parallelism and fixed SIMD width of CPUs and GPUs, on FPGAs the single work-item programming model enables pipeline parallelism for loops along with flexible amounts of data parallelism. %
In addition, FPGA local memory can be tailored to desired data widths and depths for parallel data accesses for the computation of intermediates.

Despite numerous successful applications of FPGAs as energy-efficient accelerators in many scientific domains, e.g.\ linear algebra~\cite{DeMatteis20, Gorlani19, Meyer22, Zhang22}, the current FPGA-accelerated atomistic simulations~\cite{Herbordt05, Herbordt10, Herbordt13, Herbordt19, Jones22} are mainly based on classical molecular dynamics.
Since the ERI computation is a key component for building the Hamiltonian matrix in hybrid DFT, this work paves the way for FPGA-accelerated quantum-mechanics-based atomistic simulations.
Our contributions include:

\begin{itemize}
  \item We present the first FPGA implementation targeting ERI quartets up to angular momenta of $L=3$ (256 quartet classes). The only previous attempt~\cite{ki-uf-a} was limited to the simplest quartet class with angular momentum $L=0$.
  \item Adapting to the different computation characteristics of quartet classes, 256 FPGA kernels are specifically customized in terms of parallelism and local memory layout by taking advantage of function templates using DPC++ in the Intel FPGA Add-on for oneAPI Base Toolkit~\cite{FPGAOpt4oneAPI}.
  \item A performance model is embedded into the optimization process and accurately matches the final results.
  \item This is the first implementation seamlessly integrating ERI computation and ERI compression using arbitrary bitwidth integer for the mitigation of demanding memory requirements in AMID simulations and of data transfers via PCIe from FPGA device back to host.
  \item Evaluation reveals that the FPGA kernels on 2 Stratix 10 GX 2800 cards outperform libint~\cite{libint2}, a well-established library for the ERI computation in major atomistic simulation programs~\cite{libintwiki}, parallelized on two Intel Xeon Gold 6148 CPUs (40 cores) by factors up to 6.0x and on two AMD EPYC 7713 CPUs (128 cores) by up to 1.9x.
\end{itemize}
The source code is publicly available ~\cite{zenodo}.

\section{Algorithmic Background}\label{sec:background}

The need for the computation of ERIs stems from the explicit inclusion of the electron-electron interaction Hamiltonian in hybrid DFT calculations or correlated electronic structure methods. This section describes the mathematical background of ERIs, starting from the basis functions, Cartesian GTOs, used for the representation of the wave functions. %

\subsection{Cartesian GTO}\label{subsec:gto}
A normalized primitive Cartesian GTO is defined as
\begin{IEEEeqnarray}{c}
  g_{\vb{a},\vb{R}_{A}}(\vb{r}) = N \cdot e^{-\alpha \abs{\vb{r} - \vb{R}_A}^2}\prod_{\xi \in \{x,y,z\}} \left(r_\xi - R_{A_\xi}\right)^{a_\xi} 
\end{IEEEeqnarray}
\noindent with the orbital exponent $\alpha$ and at the center $\vb{R}_A = [R_{A_x}, R_{A_y}, R_{A_z}]$.
$N$ is a normalization factor chosen so that the axis-aligned Cartesian GTOs are normalized~\cite{Kenny08}. The orbital angular momentum of $g_{\vb{a},\vb{R}_{A}}(\vb{r})$ is defined as $L_a = a_x + a_y + a_z$.
The Cartesian components of $L_a$ form an integer vector $\vb{a} = [a_x, a_y, a_z]$.
A set of all Cartesian GTOs that differ in the elements of $\vb{a}$, but share the same $L_a$, constitutes a \emph{shell} and is denoted as $a = \{ g_{\vb{a},\vb{R}_{A}}(\vb{r}) \}$.
The number of Cartesian GTOs in a shell $a$ is determined by $n_{g_a} = (L_a + 1) (L_a + 2) / 2$. The most commonly used shells in AIMD simulations are $s$, $p$, $d$, and $f$, i.e.\ $L_a = 0$, 1, 2, and 3, respectively.

\subsection{ERI Quartets}\label{subsec:bg:quartets}

An ERI describes the repulsion between the densities of two electrons, one at $\vb{r}$ and the other at $\vb{r}'$.
The density of the electron at $\vb{r}$ is the weighted sum of products of two Cartesian GTOs. Hence, an ERI is defined as
\begin{align}
  [ab|cd] = \iint \ud \vb{r} \ud \vb{r}'\,
  \frac{g_{\vb{a},\vb{R}_{A}}(\vb{r})  g_{\vb{b},\vb{R}_{B}}(\vb{r}) g_{\vb{c},\vb{R}_{C}}(\vb{r}') g_{\vb{d},\vb{R}_{D}}(\vb{r}')}
  {\abs{\vb{r} - \vb{r}'}},
  \label{eq:eri}
\end{align}
\noindent which is a 6-dimensional integral over the Cartesian coordinates of both electrons~\cite{Boys50}.
$[ab|cd]$ is the conventional notation for an ERI quartet class consisting of 4 normalized primitive Cartesian GTOs, which forms the first set of inputs to the ERI computation (Fig.~\ref{fig:fpga_design})
\begin{IEEEeqnarray}{c}
  \label{eq:input_g_abcd}
  G_{abcd} = \qty{g_{\vb{a},\vb{R}_{A}}(\vb{r}),  g_{\vb{b},\vb{R}_{B}}(\vb{r}),
                  g_{\vb{c},\vb{R}_{C}}(\vb{r}'), g_{\vb{d},\vb{R}_{D}}(\vb{r}')}.
\end{IEEEeqnarray}
Due to the eightfold permutation symmetry, an $[ab|cd]$ quartet class is mathematically identical to the variants
\begin{IEEEeqnarray}{Cl}
    & [ab|cd] = [ba|cd] = [ab|dc] = [ba|dc] \nonumber\\
  = & [cd|ab] = [cd|ba] = [dc|ab] = [dc|ba], \nonumber
\end{IEEEeqnarray}
A quartet class $[ab|cd]$ with arbitrary $a$, $b$, $c$, and $d$ is called a \emph{generic}, whereas a \emph{canonical} $[ab|cd]$ quartet class is uniquely defined by the conditions
\begin{IEEEeqnarray}{c}
  \IEEEyesnumber\label{eq:can_abcd}
  L_a \ge L_b, L_c \ge L_d, \text{ and } n_{g_a} n_{g_b} \ge n_{g_c} n_{g_d}.
\end{IEEEeqnarray}
For angular momenta up $L_a=3$ there are 55 \emph{canonical} $[ab|cd]$ quartet classes for 256 \emph{generic} quartet classes.

\subsection{ERI Computation}

\begin{figure}
\centering
\includegraphics[width=0.98\columnwidth]{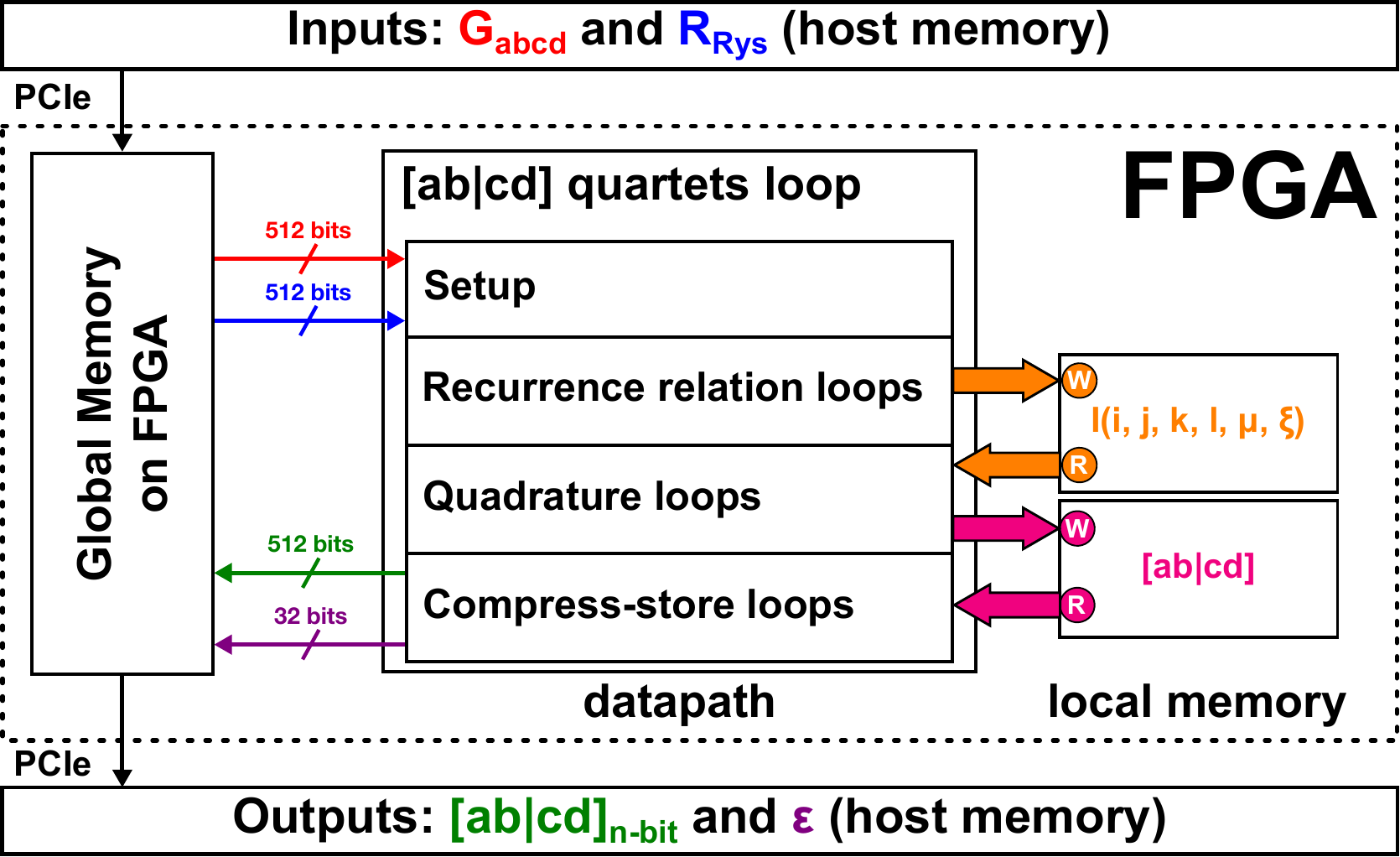}
\caption{FPGA design for ERI computation and compression.}
\label{fig:fpga_design}
\end{figure}

In the past several decades, plenty of algorithms for the ERI computation were devised~\cite{HJO_ch9}.
These algorithms can be classified into three major categories based on the mathematical formulations: Rys quadrature~\cite{dupuis1976DKR, king1976DKR, Rys83}, Obara-Saika schemes~\cite{Obara86, headgordon1988HGPRR}, and McMurchie-Davidson schemes~\cite{mcmurchie1978McMurchieDavidson}.
In the current work, we have implemented the Rys quadrature algorithm, which is favorable for FPGA acceleration because of a low memory footprint for intermediates enabling the use of fast FPGA on-chip memory to achieve parallel loads and stores. In addition, compared to the other approaches, the Rys quadrature tends to need fewer floating-point operations for the quartet classes with higher angular momenta. Due to the numerical stability of the performance relevant stages in the Rys approach, one can make use of the single-precision floating-point arithmetic operation on FPGAs\footnote{The Rys roots and weights are calculated in double-precision on the CPU in this work.}.
The flow of the computation is shown in Fig.~\ref{fig:fpga_design}.

The central principle of the Rys quadrature is to reduce the 6-dimensional integration in ERI computation (Eq.~\ref{eq:eri}) to a sum of many 1-dimensional integrations.
These 1-dimensional integrations are then solved by using Gaussian quadrature in conjunction with a set of orthonormal Rys polynomials.
The roots $t_{\mu}$ and weights $w_{\mu}$ of the Rys polynomials form the second set of input data for ERI computation (Fig.~\ref{fig:fpga_design})
\begin{IEEEeqnarray}{c}
  \label{eq:input_r_rys}
  R_{\text{Rys}} = \qty{t_{\mu}, w_{\mu} \mid \mu \in [1, n_{\text{Rys}}]},
\end{IEEEeqnarray}
\noindent where $n_{\text{Rys}}$ is the order of the Rys polynomials. %

After preparation of the inputs $G_{abcd}$ and $R_{\text{Rys}}$, our implementation of the Rys algorithm~\cite{Rys83} consists of three stages.

\subsubsection{Setup Stage}
Two auxiliary arrays $\vb{B} \in \mathbb{R}^{3 \times n_{\text{Rys}}}$ and $\vb{C} \in \mathbb{R}^{6 \times n_{\text{Rys}}}$ need to be set up using the inputs of the ERI computation and can be found in the original paper~\cite{Rys83}.

\subsubsection{Recurrence Relation Stage}
The recursive intermediates are quantities arranged in a 6-dimensional array denoted as $I(i, j, k, l, \mu, \xi)$, where the indices $i$, $j$, $k$, and $l$ are for the shells $a$, $b$, $c$, and $d$, respectively. $\mu$ enumerates the Rys polynomials and $\xi$ represents the $x$, $y$, or $z$-axis.

Starting with the initial intermediates at the ``origins'', $I(0, 0, 0, 0, \mu, \xi)$, whose numerical values depend on the $\mu$-th root and weight of the Rys polynomials and $x$, $y$, and $z$-axes, the entire $I(i, j, k, l, \mu, \xi)$ array can be built by using two distinct sets of multi-dimensional recurrence relations (RRs).

Firstly, the vertical recurrence relations (VRRs) are used, which rely on both orbital angular momenta $n$, $m$ and Cartesian axes $\xi$ of shells $a$, $b$, $c$, and $d$:
\begin{IEEEeqnarray}{rCl}
  \IEEEyesnumber\label{eq:vrrs}
                        I(i + 1, 0, k    , 0, \mu, \xi) & = &
        n B_{2, \mu}    I(i - 1, 0, k    , 0, \mu, \xi) \nonumber \\
& & +\: m B_{1, \mu}    I(i    , 0, k - 1, 0, \mu, \xi) \nonumber \\
& & +\:   C_{\xi, \mu}  I(i    , 0, k    , 0, \mu, \xi) \IEEEyessubnumber \\
                        I(i    , 0, k + 1, 0, \mu, \xi) & = &
        m B_{3, \mu}    I(i    , 0, k - 1, 0, \mu, \xi) \nonumber \\
& & +\: n B_{1, \mu}    I(i - 1, 0, k    , 0, \mu, \xi) \nonumber \\
& & +\:   C_{2\xi, \mu} I(i    , 0, k    , 0, \mu, \xi) \IEEEyessubnumber.
\end{IEEEeqnarray}
\noindent Afterwards, the horizontal recurrence relations (HRRs) are applied to build the entire $I(i, j, k, l, \mu, \xi)$ array
\begin{IEEEeqnarray}{rCl}
  \IEEEyesnumber\label{eq:hrrs}
                            I(i    , j    , k    , 0    , \mu, \xi) & = &
                            I(i + 1, j - 1, k    , 0    , \mu, \xi) \nonumber \\
& & \kern-2ex + (R_{A_{\xi}} - R_{B_{\xi}}) I(i    , j - 1, k    , 0    , \mu, \xi) \IEEEyessubnumber \\
                            I(i    , j    , k    , l    , \mu, \xi) & = &
                            I(i    , j    , k + 1, l - 1, \mu, \xi) \nonumber \\
& & \kern-2ex + (R_{C_{\xi}} - R_{D_{\xi}}) I(i    , j    , k    , l - 1, \mu, \xi).\IEEEyessubnumber
\end{IEEEeqnarray}
\noindent Unlike VRRs, the HRRs only depend on the Cartesian coordinates of the 4 shells in $[ab|cd]$.
These RRs can be implemented with iterative loops, but all of them contain read-after-write (RAW) dependencies on previous iterations.

\subsubsection{Quadrature Stage}
All ERIs of an $[ab|cd]$ quartet class are computed by Gaussian quadrature
\begin{IEEEeqnarray}{c}
  [ab|cd] = \sum_{\mu = 1}^{n_{\text{Rys}}} w_{\mu} \left[\prod_{\xi \in \{x, y, z\}} I(a_{\xi}, b_{\xi}, c_{\xi}, d_{\xi}, \mu, \xi)\right],
  \label{eq:gauq}
\end{IEEEeqnarray}
\noindent where the indices $a_{\xi}$, $b_{\xi}$, $c_{\xi}$, and $d_{\xi}$ are the angular momenta components of the GTOs $a$, $b$, $c$, and $d$, respectively.

\subsection{ERI Compression}

A real-world AIMD simulation may need to compute several billions up to trillions of ERIs in each time step~\cite{Kuehne2020}.
In order to reduce the high demand on memory usage, several different algorithms for ERI compression were developed in the past~\cite{Fülscher93, Guidon08, Ying15}.
In this work, we adapt the ERI compression algorithm of Guidon et~al.~\cite{Guidon08} which is also used by the quantum chemistry software package CP2K~\cite{Kuehne2020} and directly integrated with the ERI calculation as fourth stage to make use of FPGA pipelining.

After computing all ERIs for one $[ab|cd]$ quartet, the maximum absolute value is denoted as $b_{\text{max}}$.
Targeting an $n$-bit encoding, the maximum representable signed integer is $2^{n-1}-1$ and the ``quantum value'' for representing the ERIs of $[ab|cd]$ is
\begin{IEEEeqnarray}{c}
  \epsilon = b_{\text{max}} \cdot ({2^{n - 1} - 1})^{-1}.
  \label{eq:calc_epsilon}
\end{IEEEeqnarray}
Then the compressed ERIs in $n$-bit signed integers can be represented as multiples of the quantum value
\begin{IEEEeqnarray}{c}
  [ab|cd]_{n\text{-bit}} = \text{ANINT}([ab|cd]\cdot\epsilon^{-1}),
\end{IEEEeqnarray}
\noindent where the function ANINT returns the nearest integer number of its argument.
An $[ab|cd]_{n\text{-bit}}$ array together with the scaling factor $\epsilon$ form the outputs of combined ERI computation and compression for one ERI quartet (Fig.\ref{fig:fpga_design}).

During an AIMD simulation, $[ab|cd]_{n\text{-bit}}$ will be decompressed to floating-point numbers on the host as
\begin{IEEEeqnarray}{c}
  \widetilde{[ab|cd]} = [ab|cd]_{n\text{-bit}} \cdot \epsilon.
  \label{eq:decomp_eris}
\end{IEEEeqnarray}
\noindent It can be shown that the maximum absolute error between original $[ab|cd]$ and $\widetilde{[ab|cd]}$ is bound by $\epsilon / 2$ for a quartet, because the rounding error of ANINT is $\pm 1/2$.

\section{Design and Implementation}\label{sec:design}

\begin{figure*}
\centering
\includegraphics[width=0.98\textwidth]{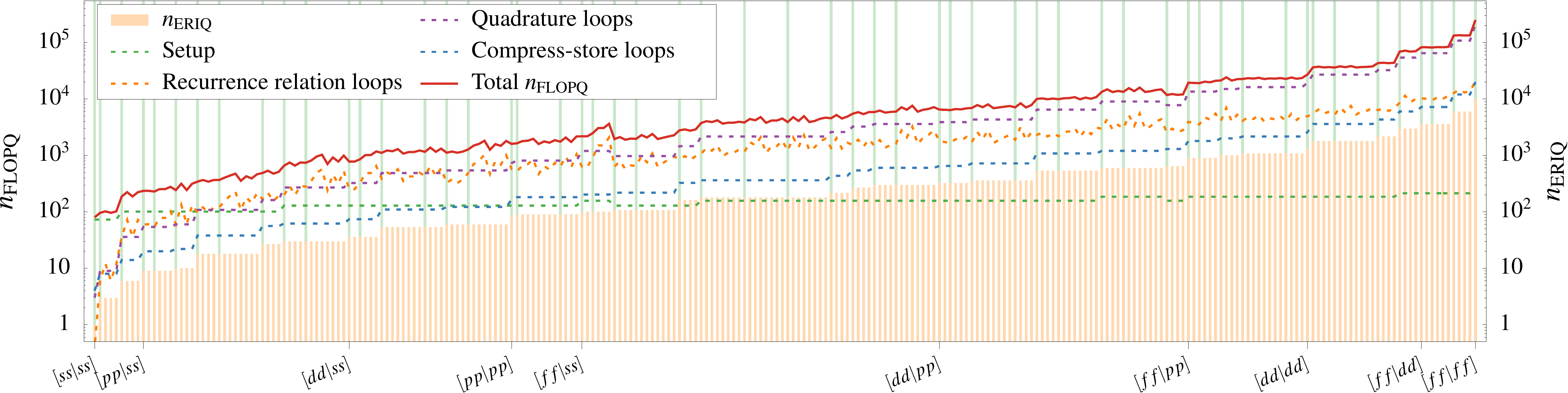}
\caption{Number of ERIs per quartet ($n_{\text{ERIQ}}$, vertical bars), number of floating-point operations per quartet ($n_{\text{FLOPQ}}$) for the individual stages (dotted lines), and total $n_{\text{FLOPQ}}$ (solid red line) for 256 \emph{generic} ERI quartet classes.
55 \emph{canonical} quartet classes are highlighted in green bars.
Only the \emph{canonical} ERI quartet classes of form $[aa|cc]$ are explicitly labeled for clarity.}
\label{fig:num_FLOPs_256}
\end{figure*}

In this section, we present how the introduced algorithm is realized on FPGA. Fig.~\ref{fig:fpga_design} is complemented with an overview of the loop structure in Fig.~\ref{fig:fpga_algorithm}.
The perspective herein is often presented in terms of loop structures.
This view is a guide for the high-level synthesis tools to generate efficient hardware structures, in particular parallel and pipelined datapaths and matching layouts of local memory resources.
Based on the general structure in Fig.~\ref{fig:fpga_design} and Fig.~\ref{fig:fpga_algorithm}, a customized design is created for each ERI quartet class. %
Single-precision floating-point arithmetic is utilized in the FPGA implementation because of numerical stability of the Rys quadrature.
Before discussing the detailed design choices, we analyze the computational requirements of different quartet classes.

\subsection{Analysis of ERI Quartet Classes}
\label{subsec:function_template}

Since each shell in $[ab|cd]$ may be any one of $s$, $p$, $d$, and $f$ GTOs (Section \ref{subsec:gto}), there are $4^4=256$ \emph{generic} ERI quartet classes that possess dramatically different computation characteristics regarding the number of floating-point operations per quartet and the number of required intermediates.
Fig.~\ref{fig:num_FLOPs_256} presents an overview of the number of ERIs per quartet ($n_{\text{ERIQ}}$) together with the number of required FLOPs ($n_{\text{FLOPQ}}$), individually for the four stages indicated in Fig.~\ref{fig:fpga_design} and~\ref{fig:fpga_algorithm}, i.e.\ \emph{setup}, \emph{recurrence relation loops}, \emph{quadrature loops}, and \emph{compress-store loops}, as well as summed up as total $n_{\text{FLOPQ}}$ for the 256 \emph{generic} quartet classes, which are sorted by means of ascending $n_{\text{ERIQ}}$ from left to right, e.g.\ starting from $[ss|ss]$ with merely 1 ERI per quartet, to the rightmost $[ff|ff]$ with $n_{\text{ERIQ}} = 10000$.
$n_{\text{FLOPQ}}$ is obtained by counting the addition, multiplication, and division operations for each $[ab|cd]$ quartet.\footnote{%
There are 6 - 18 division operations in the setup stage and 1 division while computing $\epsilon^{-1}$ for ERI compression.
By considering the fact that total $n_{\text{FLOPQ}}$ is dominated by $10^2$ - $10^5$ addition and multiplication operations, division is counted as one operation for simplicity.}
The total $n_{\text{FLOPQ}}$ (solid red line in Fig.~\ref{fig:num_FLOPs_256}) increases drastically in line with $n_{\text{ERIQ}}$, e.g.\ from 80 for $[ss|ss]$ ($n_{\text{ERIQ}} = 1$) to $2.5 \times 10^{5}$ for $[ff|ff]$ ($n_{\text{ERIQ}} = 10^{4}$).
The workload ratios of the four algorithmic stages also differ a lot between different $[ab|cd]$ quartet classes, with a few small quartet classes being dominated by the setup stage, intermediate classes requiring most work in the recurrence relation and quadrature stages, and for the larger classes quadrature requiring far more operations than other stages. Adaptation to this imbalance, along with requirements regarding local memory layout and data dependencies, is a central reason for creating customized designs for each quartet class.

\subsection{Templated FPGA Kernel Architecture}\label{subsec:kernel_architecture}

To create designs customized for different quartet classes we leverage C++ function template parameters that are supported by the Intel FPGA Add-on for oneAPI Base Toolkit~\cite{FPGAOpt4oneAPI} %
and by the SYCL standard it builds upon. %
The orbital angular momenta in the shells of an $[ab|cd]$ quartet class, $L_a$, $L_b$, $L_c$, and $L_d$, are passed as template parameters to the FPGA kernels.
Other derived parameters, $n_{g_a}$, $n_{g_b}$, $n_{g_c}$, $n_{g_d}$, and $n_{\text{Rys}}$, are used as constants evaluated at compile-time. Thereafter, customized designs for 256 ERI quartet classes were synthesized separately into individual bitstreams to assess the best performance for each quartet class.

The benefits of the template parameters are the customization of local memory sizes, which allows to create parallel data paths from loops with trip counts known at compile-time via unrolling annotations.
Such unrolling is indicated in Fig.~\ref{fig:fpga_algorithm} with \textbf{for all} statements in lines \ref{lst:line:forall_rr} and \ref{lst:line:forall_quad} for the \emph{recurrence relation} and \emph{quadrature loops}, but also performed within \emph{setup} and inner blocks of the \emph{compress-store loops} (not elaborated in Fig.~\ref{fig:fpga_algorithm}). However, the customization based on the template parameters goes further in two regards. Firstly, based on a model of expected throughput, further loops can be unrolled when they would otherwise become a performance bottleneck. Secondly, the layout of intermediates in local memory is adapted to match the loop parallelism. In practice, the optimization strategy combines an analytic performance model with empirically found heuristics for resource consumption and routability of different design points.
In the following subsections, we present this strategy for optimizing the throughput for an individual quartet as governed by the trip counts of the four stages. Note that to realize the full potential of the design, these stages also need to overlap during the computation of multiple different quartets within the same class (Fig.~\ref{fig:fpga_algorithm} line~\ref{lst:line:quartets}), which will be discussed in Sec.~\ref{subsec:quartet_loop}.

\begin{figure}[tbh]
\begin{algorithmic}[1]
\REQUIRE $G_{abcd}$, $R_{\text{Rys}}$
\ENSURE compressed\_quartets $[ab|cd]_{n\text{-bit}}$, $\epsilon$
  \STATE \emph{$[ab|cd]$ quartets loop}:\label{lst:line:quartets}
  \FOR{$quartet \in$ quartet class $[ab|cd]$}
  \STATE \emph{Setup}:
  \STATE build auxiliary arrays $\vb{B_{\text{FF}}}$ and $\vb{C_{\text{FF}}}$
  \STATE \emph{Recurrence relation loops}:
  \FOR{Cartesian axis $\xi \in \{x, y, z\}$}
    \FOR{$\mu$-th Rys polynomial $\in [1, n_\text{Rys}]$}
      \STATE calculate $I_{\text{FF}}(i, 0, k, 0)$ via VRRs\label{lst:line:vrr}
      \STATE calculate $I_{\text{FF}}(i, j, k, l)$ via HRRs\label{lst:line:hrr}
      \FOR{$l \in [1, L_d]$}\label{lst:line:for_rr}
        \FORALL{$k \in [1, L_c]$, $j \in [1, L_b]$, $i \in [1, L_a]$}\label{lst:line:forall_rr}
            \STATE $I(i, j, k, l, \mu, \xi) \leftarrow I_{\text{FF}}(i, j, k, l)$
        \ENDFOR
      \ENDFOR
    \ENDFOR
  \ENDFOR
  \STATE \emph{Quadrature loops}:
  \FOR{GTOs in shell $d \in [1, n_{g_d}]$}\label{lst:line:for_quad}
    \FOR{GTOs in shell $c \in [1, n_{g_c}]$}
        \FORALL{$b \in [1, n_{g_b}]$, $a \in [1, n_{g_a}]$}\label{lst:line:forall_quad}
            \STATE calculate $[ab|cd]$ integral
            \STATE $b_{\text{max}} \leftarrow \max(b_{\text{max}}, \abs{[ab|cd]})$
        \ENDFOR
    \ENDFOR
  \ENDFOR
  \STATE $\epsilon \leftarrow b_{\text{max}} \cdot ({2^{n - 1} - 1})^{-1}$
  \STATE \emph{Compress-store loops}:\label{lst:line:csloop}
  \STATE $\vb{Y}_{\text{FF}}^{\text{rem}} \leftarrow {0}$
  \FOR{$chunk \in [1, n_{\text{CS}}]$}
    \IF{content($\vb{Y}_{\text{FF}}^{\text{rem}}$) $<$ 512-bit}
        \STATE $\vb{X}_{\text{FF}}^{ab} \leftarrow \text{ANINT}([**|cd]\cdot\epsilon^{-1})$
    \ENDIF
    \STATE $\vb{Z}_{\text{FF}}^{\text{512-bit}} \leftarrow \vb{X}_{\text{FF}}^{ab} \text{ and/or } \vb{Y}_{\text{FF}}^{\text{rem}}$
    \STATE compressed\_quartets[$chunk$] $\leftarrow \vb{Z}_{\text{FF}}^{\text{512-bit}}$
    \STATE $\vb{Y}_{\text{FF}}^{\text{rem}} \leftarrow \text{remainder of } \vb{X}_{\text{FF}}^{ab}$
  \ENDFOR
  \ENDFOR
\end{algorithmic}
\caption{Loop structure of FPGA kernels. \textbf{for all} loops are fully unrolled for all designs, \textbf{for} loops can be pipelined loops or fully unrolled loops depending on the quartet class.}
\label{fig:fpga_algorithm}
\end{figure}

\subsection{General Optimization and Local Memory Layout}\label{sec:general_optimization}

Conceptually, the throughput optimization starts from the constraints imposed by the external memory interface. The Intel Stratix 10 GX 2800 FPGA in a Bittware 520N card has four physical DDR4 memory channels. Here, these memory channels are used in interleaved mode, creating a 512-bit memory interface to the kernels that can supply data at kernel clock rates (as opposed to a 300 MHz throughput limit without interleaving).
The input data consist of $G_{abcd}$ (Eq.~\ref{eq:input_g_abcd}) and $R_{\text{Rys}}$ (Eq.~\ref{eq:input_r_rys}, which are encoded in two 512-bit words for every quartet, such that asymptotically two cycles per quartet will be spent on the inputs. Similarly, a constant size for global memory write per quartet is needed for the quantum value, $\epsilon$ (32 bits).
Consequently, for all but the smallest quartet classes, the off-chip memory access is dominated by the output of $[ab|cd]_{n\text{-bit}}$, the compressed ERIs represented by $n$-bit signed integer. In 16-bit compression format, $512/16=32$ ERIs can be written out per 512-bit word, or more generally in $n$-bit format $\lfloor512/n\rfloor$ ERIs per chunk after padding can be written out in 
$n_{\text{CS}} = \left\lceil \frac{n_{\text{ERIQ}}}{\left\lfloor 512/n \right\rfloor} \right\rceil$ 
cycles. These values define the target throughput of the phases before the \emph{compress-store loops}. Unless explicitly stated otherwise, in the remainder of the text, we use the 16-bit encoding and 32 ERIs per cycle target for illustrative purposes, whereas in the actual implementation, these are part of the compile-time optimizations based on template parameters.
To match the throughput of the \emph{compress-store loops}, the trip counts of iterative pipelined loops of the preceeding phases should not exceed $n_{\text{CS}}$. Now discussing the phases in their normal sequence, we approached this goal as follows:

\subsubsection{Setup}\label{paragraph:default:setup}

The \emph{setup} stage builds two auxiliary arrays $\vb{B} \in \mathbb{R}^{3 \times n_{\text{Rys}}}$ and $\vb{C} \in \mathbb{R}^{6 \times n_{\text{Rys}}}$.
With $n_{\text{Rys}} \in [1, 8)$, depending on the $[ab|cd]$ quartet, both arrays are small and thus fit into FPGA registers (denoted as $\vb{B}_{\text{FF}}$ and $\vb{C}_{\text{FF}}$ in Fig.~\ref{fig:fpga_algorithm}).
Since all involved loops have low trip counts, i.e.\ 3 for Cartesian axes and $n_{\text{Rys}}$, for simplicity all loops are fully unrolled. %

\subsubsection{Recurrence Relation Loops}\label{paragraph:default:rrloops}

\begin{table*}
\renewcommand{\arraystretch}{1.3}
\caption{Default layout for $I(i, j, k, l, \mu, \xi)$ in FPGA local memory. Note that actual dimensions always depend on the template parameters.}
\label{tab:intermediate_I}
\centering
\begin{tabular}{c|c|c|c|>{\centering\arraybackslash\hspace{0pt}}m{3.5cm}|>{\centering\arraybackslash\hspace{0pt}}m{3.5cm}}
\hline
Index & Interval & Padded length$^{1}$ & Memory layout &
Write for $ijk$-unrolled recurrence relation loops & Read for $\xi\mu ab$-unrolled quadrature loops \\
\hline
$\mu$ & $[1, n_{\text{Rys}}]$ & $\lceil n_{\text{Rys}}\rceil_{\text{pow2}}$ & bank width & sequential & parallel   \\
$\xi$ & $[1, 3]$              & $4$                                         & banks      & sequential & parallel   \\
$i$   & $[0, L_a + L_b]$      & $\lceil L_a + L_b + 1 \rceil_{\text{pow2}}$ & banks      & parallel   & parallel   \\
$j$   & $[0, L_b      ]$      & $\lceil L_b +       1 \rceil_{\text{pow2}}$ & banks      & parallel   & parallel   \\
$k$   & $[0, L_c + L_d]$      & $\lceil L_c + L_d + 1 \rceil_{\text{pow2}}$ & banks      & parallel   & sequential \\
$l$   & $[0, L_d      ]$      & not padded                                  & -          & sequential & sequential \\
\hline
\multicolumn{6}{l}{\footnotesize$^1\,\lceil n \rceil_{\text{pow2}}$ returns the smallest integer of power of 2 and also $\ge n$.}
\end{tabular}
\end{table*}

In this phase, first, a small buffer that fits into FPGA registers, $I_{\text{FF}}(i, j, k, l)$ in Fig.~\ref{fig:fpga_algorithm}, is employed for the computations of VRRs and HRRs (lines~\ref{lst:line:vrr}, \ref{lst:line:hrr}) and the involved loops are fully unrolled.
This allows the compiler tools to resolve the inherent data dependencies by creating a sufficiently deep datapath that computes a full $I_{\text{FF}}(i, j, k, l)$ set per cycle, albeit with a certain latency.
$\xi \times \mu $ sets $I_{\text{FF}}(i, j, k, l)$ are computed in this way and buffered in $I(i, j, k, l, \mu, \xi)$ to form the input required by the next phase. In the default configuration as depicted in Fig.~\ref{fig:fpga_algorithm}, this is in parallel for the inner $ijk$ loops that are unrolled, but sequential for the $\xi$, $\mu$, and $l$ loops. Such approach is often fast enough for the overall throughput goal and helps to find a suitable local memory layout for $I(i, j, k, l, \mu, \xi)$ that works for all quartet classes.

The size of $I(i, j, k, l, \mu, \xi)$ is larger than $I_{\text{FF}}(i, j, k, l)$ and it grows steeply with angular momenta of $[ab|cd]$, e.g.\ only 3 for $[ss|ss]$, but 5376 for $[ff|ff]$.
Thus, except for small quartet classes, such a buffer is generally best implemented in on-chip memory resources, i.e.\ block RAMs or MLABs.
In order to achieve stall-free local memory access to $I(i, j, k, l, \mu, \xi)$, the bank layout of FPGA local memory must be consistent with the corresponding unrolled loop structure both of the writing \emph{recurrence relation loops} and of the using \emph{quadrature loops}, which requires parallel access into the $\mu\xi$ dimensions, as already visible in Eq.~\ref{eq:gauq}.
The default layout
for $I(i, j, k, l, \mu, \xi)$ that fulfills both requirements is illustrated in Table~\ref{tab:intermediate_I} along with the access patterns from both loops.
The bank width of local memory is configured in the $\mu$ dimension with padded length of $n_{\text{Rys}}$ to the power of 2.
The number of memory banks is defined by the multiplied padded lengths of $\xi ijk$ dimensions.
This leaves only the last dimension $l$ as memory depth with sequential access.
With this customized banking geometry and $ijk$-unrolled \emph{recurrence relation loops}, the stores from $I_{\text{FF}}(i, j, k, l)$ to the $I(i, j, k, l, \mu, \xi)$ array are stall-free and parallel in $ijk$ dimensions.
The write operations in other dimensions, however, are sequential with a trip count of $n_{\text{RR}} = 3 n_{\text{Rys}} (L_d + 1)$. %
Lastly, these nested sequential loops are coalesced by compiler directive to reduce the FPGA area overhead.

\subsubsection{Quadrature Loops}\label{paragraph:default:gqloops}

\begin{table*}
\renewcommand{\arraystretch}{1.3}
\caption{Default layout for $[ab|cd]$ in FPGA local memory.}
\label{tab:intermediate_abcd}
\centering
\begin{tabular}{c|c|c|c|>{\centering\arraybackslash\hspace{0pt}}m{3.0cm}|>{\centering\arraybackslash\hspace{0pt}}m{3.7cm}}
\hline
Indices & Interval & Padded length$^{1}$ & Memory layout &
Write for $\xi\mu ab$-unrolled quadrature loops & Read for $ab$-unrolled compress-store loop \\
\hline
$a$ and $b$ & $a \in [1, n_{g_a}]$ and $b \in [1, n_{g_b}]$ & $\lceil n_{g_a} n_{g_b}\rceil_{\text{pow2}}$ & banks & parallel   & parallel \\
$c$ and $d$ & $c \in [1, n_{g_c}]$ and $d \in [1, n_{g_d}]$ & not padded & - & sequential & sequential \\
\hline
\multicolumn{6}{l}{\footnotesize$^1\,\lceil n \rceil_{\text{pow2}}$ returns the smallest integer of power of 2 and also $\ge n$.}
\end{tabular}
\end{table*}

The \emph{quadrature loops} read $I(i, j, k, l, \mu, \xi)$ from the preceding phase and calculate all ERIs of an $[ab|cd]$ quartet as output.%
The quadrature stage is composed of 6-fold nested loops, in which the 4 outer loops run over each dimension of $[ab|cd]$ and 2 inner loops perform the multiplications and additions for $\xi$ and $\mu$, respectively (Eq.~\ref{eq:gauq}), which are always unrolled without data dependencies.
With the further parallel execution in the $a$ and $b$ dimensions as indicated in Fig.~\ref{fig:fpga_algorithm} and Tab.~\ref{tab:intermediate_I}, the remaining sequential loops have a trip count of $n_{\text{GQ}} = n_{g_d} n_{g_c}$.

With this structure, $n_{g_a} n_{g_b}$ ERIs, denoted as $[**|cd]$ in Fig.~\ref{fig:fpga_algorithm}, are generated as output of the \emph{quadrature loops} in every clock cycle and stored in the local memory.
In the FPGA kernels, the $[ab|cd]$ quartet intermediate is implemented as 2-D array using local memory.
As for this buffer, the loop patterns of the producing and the consuming loops could be matched, the memory layout as presented in Tab.~\ref{tab:intermediate_abcd} directly fits for both sides.
The $ab$ dimensions in the $[ab|cd]$ quartet buffer are fused into a single dimension with padded length of power of 2 for $n_{g_a} n_{g_b}$ as local memory banks for parallel access.
The $cd$ dimensions are also fused into the memory depth and the related nested loops of the quadrature phase are coalesced via compiler directive.
At last, the \emph{quadrature loops} also identify $b_{\text{max}}$ per quartet that is used for scaling in the next phase. This involves a parallel reduction over the unrolled loops.

\subsubsection{Compress-Store Loops}\label{paragraph:default::csloops}

The \emph{compress-store loops} (Fig.~\ref{fig:fpga_algorithm} lines~\ref{lst:line:csloop}) load $[ab|cd]$ from 
the intermediate buffer, scale individual ERIs as $n$-bit signed integer multiples of $\epsilon$, and store the compressed $[ab|cd]_{n\text{-bit}}$ into FPGA global memory, which are transferred to the host memory via PCIe in the end.

The main implementation idea of the \emph{compress-store loops} also revolves around the data layout adaption from ERIs available in parallel $a$ and $b$ dimensions (Tab.~\ref{tab:intermediate_abcd}) from the previous phase, to the output in chunks of 512 bits. The sequential part of these loops iterates over the chunks that have to be written. Since the overall sequence of ERIs computed and written is identical, the layout transformation mainly relies on a small buffer in registers $\vb{Y}_{\text{FF}}^{\text{rem}}$, that retains the ERIs that didn't fit into the previous chunk for writing them out first in the next chunk.
Initially, after loading $[**|cd]$ via local memory banks, the fetched ERIs are compressed through unrolled loops in parallel and the results are stored in $\vb{X}_{\text{FF}}^{ab}$.
When $\vb{X}_{\text{FF}}^{ab} \ge$ 512-bit, a first complete 512-bit chunk is copied to $\vb{Z}_{\text{FF}}^{\text{512-bit}}$, which has a fixed size of 512-bit.
As essential operation for each iteration of the iterative \emph{compress-store loop}, the contents in $\vb{Z}_{\text{FF}}^{\text{512-bit}}$ are written to the output $[**|cd]_{n\text{-bit}}$ in FPGA global memory.
The remaining compressed bits in $\vb{X}_{\text{FF}}^{ab}$ moved to $\vb{Y}_{\text{FF}}^{\text{rem}}$ for the next iteration.
Repeated iterations of the loop can produce an output without loading new inputs to $\vb{X}_{\text{FF}}^{ab}$
until the remainder in $\vb{X}_{\text{FF}}^{ab}$ is less than 512-bit, or until the last (possibly incomplete) 512-bit chunk is written.

\subsubsection{Initial Performance Model}\label{paragraph:default::initial_perf}

\begin{table}[tbh]
\renewcommand{\arraystretch}{1.3}
\caption{Throughput model of selected quartet classes after general optimization (left, Sec.~\ref{sec:general_optimization}) and applying further unrolling (right, Sec.~\ref{sec:further_unrolling}). Bold numbers highlight bottlenecks.}
\label{tab:initial_perf}
\centering
\begin{tabular}{c|rrr|rrr}
\hline
 & \multicolumn{3}{c|}{General Optimization}  & \multicolumn{3}{c}{Further Unrolling} \\
Quartet & $n_{\text{RR}}$ & $n_{\text{GQ}}$ & $n_{\text{CS}}$ & $n_{\text{RR}}$ & $n_{\text{GQ}}$ & $n_{\text{CS}}$ \\
\hline
$[ss|ss]$ & \textbf{3} & 1 & 1 & 1 & 1 & 1\\
$[pp|pp]$ & \textbf{18} & 9 & 3 & 3 & 3 & 3\\
$[dd|ps]$ & \textbf{9} & 3 & 4 & 3 & 3 & 4\\
$[dd|dd]$ & \textbf{45} & 36 & 41 & \textbf{45} & 36 & 41\\
$[ff|fd]$ & 54 & 60 & \textbf{188} & 54 & 60 & \textbf{188}\\
$[ff|ff]$ & 84 & 100 & \textbf{313} & 84 & 100 & \textbf{313}\\
\hline
\end{tabular}
\end{table}

A first performance model, based on the trip counts of sequential loops as discussed in this section is shown for a few selected quartet classes in Tab.~\ref{tab:initial_perf}. It shows that for classes with high angular momentum and lots of parallelism in the dimensions that are already unrolled, the \emph{compress-store loops} already form the expected bottleneck. However, for smaller quartet classes, the parallelism is not sufficient to saturate the memory interface.
Additionally, from a computational perspective, ERI compression with lower bitwidth (not illustrated in Tab.~\ref{tab:initial_perf}) can allow for even higher throughput, even though in practice the trade-off between numerical accuracy and performance eventually becomes critical. In any case, it can be seen that there is a demand to increase the parallelism particularly for smaller quartet classes, which we discuss next.

\subsection{Parameter Guided Further Unrolling at Compile-Time}\label{sec:further_unrolling}

When the throughput of the \emph{recurrence relation loops} becomes the bottleneck, as for the first four examples in Tab.~\ref{tab:initial_perf}, additional parallelism can be obtained by additionally unrolling the previously sequential loops over $l$, $\mu$, and $\xi$. Based on a model for the expected trip counts that is embedded into the source code and evaluated at compile-time in the form of \texttt{constexpr} statements, such additional unrolling is performed automatically. Tab.~\ref{tab:unroll_rr_loop} (upper half) shows how this process improves the trip count of this phase up to a fully parallel datapath. Along with the changes, the storage for $I(i, j, k, l, \mu, \xi)$ needs to be switched from BRAM to registers at the indicated transition point. An additional empirically found constraint is evaluated at compile-time to limit the basic size of intermediates in registers to 108 elements, as we saw routing or timing problems otherwise.
\begin{table}[tbh]
\renewcommand{\arraystretch}{1.3}
\caption{Strategy for further loop unrolling.}
\label{tab:unroll_rr_loop}
\centering
\begin{tabular}{ccc}
\hline
\multicolumn{3}{c}{\emph{recurrence relation loops}} \\
\hline
$I(i, j, k, l, \mu, \xi)$ & Unrolled indices & $n_{\text{RR}}$ [cycles] \\
\hline
BRAMs & $ijk$        & $3n_{\text{Rys}}(L_d + 1)$ \\
BRAMs & $ijkl$       & $3n_{\text{Rys}}$          \\
Registers     & $ijkl\mu$    & $3$                        \\
Registers     & $ijkl\mu\xi$ & $1$                        \\
\hline
\multicolumn{3}{c}{\emph{quadrature loops}} \\
\hline
$[ab|cd]$ & Unrolled indices & $n_{\text{GQ}}$ [cycles] \\
\hline
BRAMs & $\xi\mu ab$   & $n_{g_d}n_{g_c}$ \\
Registers     & $\xi\mu abc$  & $n_{g_d}$        \\
Registers     & $\xi\mu abcd$ & $1$              \\
\hline
\end{tabular}
\end{table}

After removing the $n_{\text{RR}}$ bottleneck for $[pp|pp]$ in Tab.~\ref{tab:initial_perf} -- and for other quartets already in the general case -- the \emph{quadrature loops} can also be the throughput bottleneck in the same way. Consequently, the same compile-time strategy for additional parallelism is applied here, as added in the lower half of Tab.~\ref{tab:unroll_rr_loop}. There are 12 possible combinations of the unrolling patterns of the \emph{recurrence relation} and \emph{quadrature} phases. %
Out of these, 7 are actually chosen during the compile-time optimization process as best parallelism structure for at least one quartet class. The right side of Tab.~\ref{tab:initial_perf} illustrates selected outcomes of this further unrolling, with $[dd|dd]$ representing an example where local memory implementation constraints prevent the design from matching the throughput target of the \emph{compress-store loops}.

\subsection{Concurrency in Outer Loop over Multiple Quartets}\label{subsec:quartet_loop}

Returning to a high-level perspective on the kernel architecture, it is composed of two layers of nested loops: the outermost \emph{$[ab|cd]$ quartets loop} and all other loops comprising the four stages for computing the compressed $[ab|cd]_{n\text{-bit}}$ quartet.
After the optimizations described so far, the throughput of the four individual stages is optimized. However, since each of them still has a latency that can exceed the trip count of its sequential part by far, pipelining over the outermost loop is crucial for overall good occupancy of the design. To this end, the intermediate buffers $I(i, j, k, l, \mu, \xi)$ and $[ab|cd]$ implemented in BRAM need additional capacity to hold the values for different quartets that are currently in the pipeline of the outer loop. In the Intel FPGA reports, these extra slots in the buffers are denoted as \emph{private copies} and can be generated via a $\mathtt{max\_concurrency}$ attribute. When the depth of the local memory without private copies is below the 512 entry depth typically required to fill a single RAM block, additional private copies can be obtained virtually for free, until the 512 entry threshold is reached. This is possible as no additional access ports are required, because each pipeline slot in the datapath of the inner loops can be occupied by exactly one quartet instance from the outer loop at a time.

In the general layout of local memory as presented in Tab.~\ref{tab:intermediate_I} and~\ref{tab:intermediate_abcd}, the original depths is typically low, corresponding to only one or two dimensions with small compile-time determined indices. Therefore, a high number of private copies is feasible for these designs.
As the default heuristics of the compilation tools generate far fewer private copies than required for the presented designs, we explicitly set the $\mathtt{max\_concurrency(}c_{\text{max}}\mathtt{)}$ for the entire \emph{$[ab|cd]$ quartets loop}.
The optimal value $c_{\text{max}}$ is determined empirically for each quartet class by starting from a small value, i.e.\ $c_{\text{max}} = 8$ and doubling till the measured performance saturates.

\section{Evaluation}\label{sec:evaluation}
Based on the FPGA design and implementation described in Section~\ref{sec:design}, the FPGA kernel is instantiated by using C++ function templates, and compiled and synthesized for 256 \emph{generic} $[ab|cd]$ quartet classes with the Intel FPGA Add-on for oneAPI Base Toolkit (version 22.3.0) targeting Bittware 520N board with Intel Stratix 10 GX 2800 FPGA.

\subsection{Benchmark}\label{subsec:benchmark}

As benchmark, we use a synthetic molecular system composed of 32 sites arranged on a cubic $4 \times 4 \times 2$ lattice with a lattice parameter of 1~\AA. Each site has primitive Cartesian $s$, $p$, $d$, and $f$ GTO-shells with an exponent of 1.5. No screening of quartets is used.
The performance metric is the measured FPGA kernel throughput in terms of compressed ERIs in Giga ($10^{9}$) ERIs per second (GERIS).
This evaluation ignores the preparation on the host (computation of Rys roots and weights) and the transfer of inputs and outputs between FPGA and host memory via PCIe ($\approx$ 6~GB/s), which becomes a practical bottleneck for intermediate to large quartet classes with 3 and 4 GERIS per Bittware 520N card for 16-bit and 12-bit compression, respectively.

\subsection{Synthesis Results and Throughput Analysis}\label{subsec:impl_results}

\begin{figure}[tbh]
\centering
\subfloat[ALMs: 933120]{%
  \includegraphics[width=0.46\columnwidth]{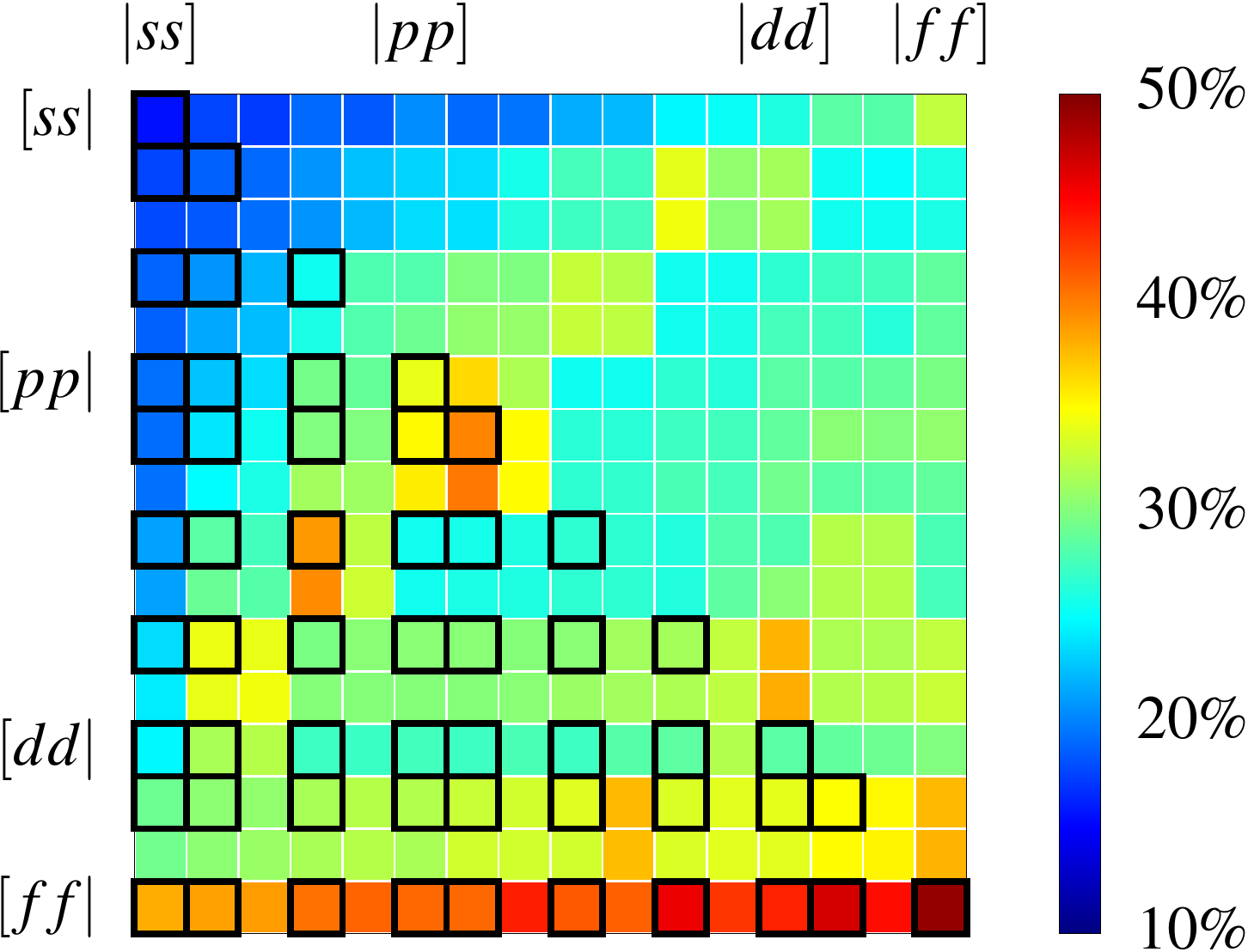}%
\label{fig:fpga_alms}}
\hfil
\subfloat[Registers: 3732480]{%
  \includegraphics[width=0.46\columnwidth]{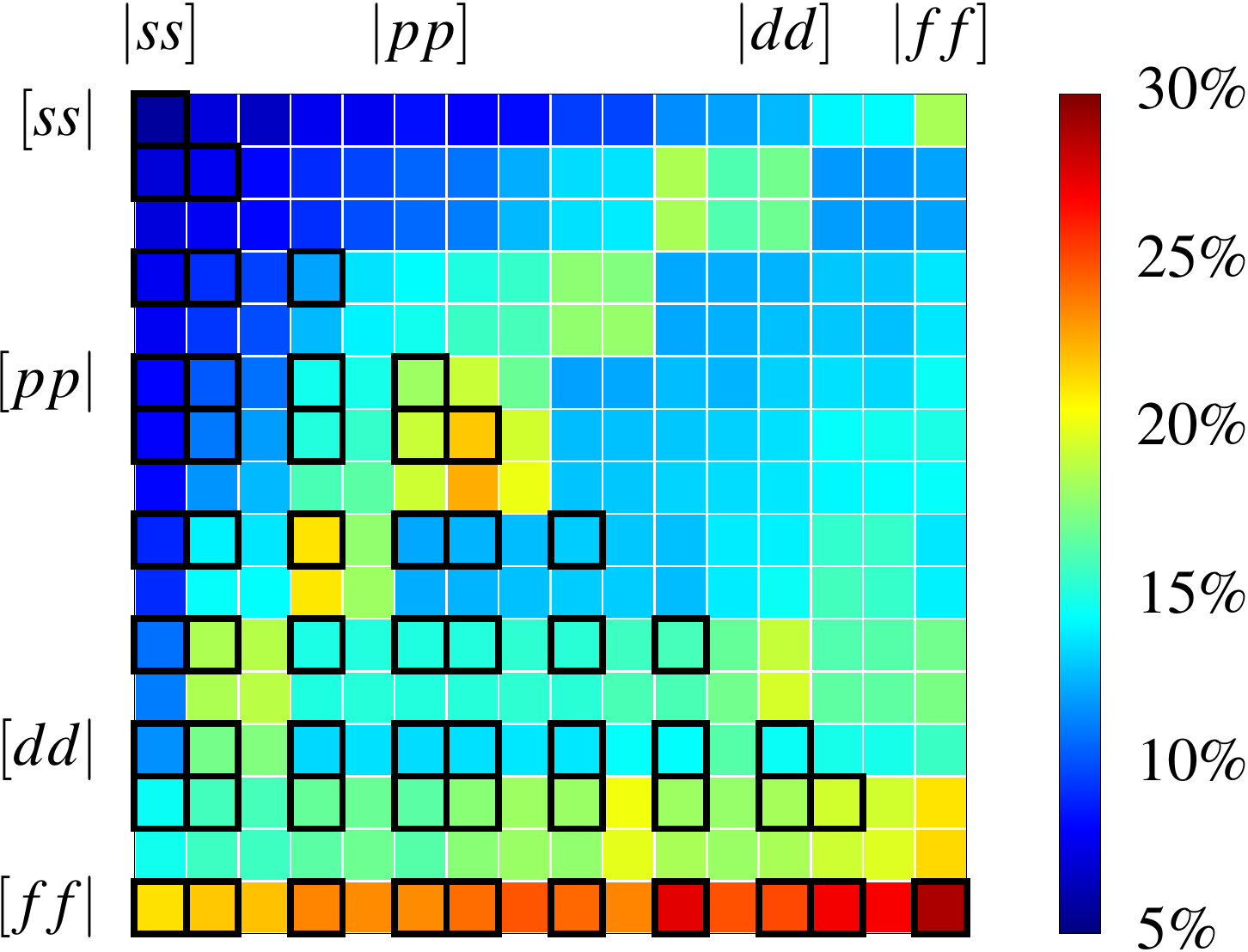}%
\label{fig:fpga_regs}}
\newline
\subfloat[DSP blocks: 5760]{%
  \includegraphics[width=0.46\columnwidth]{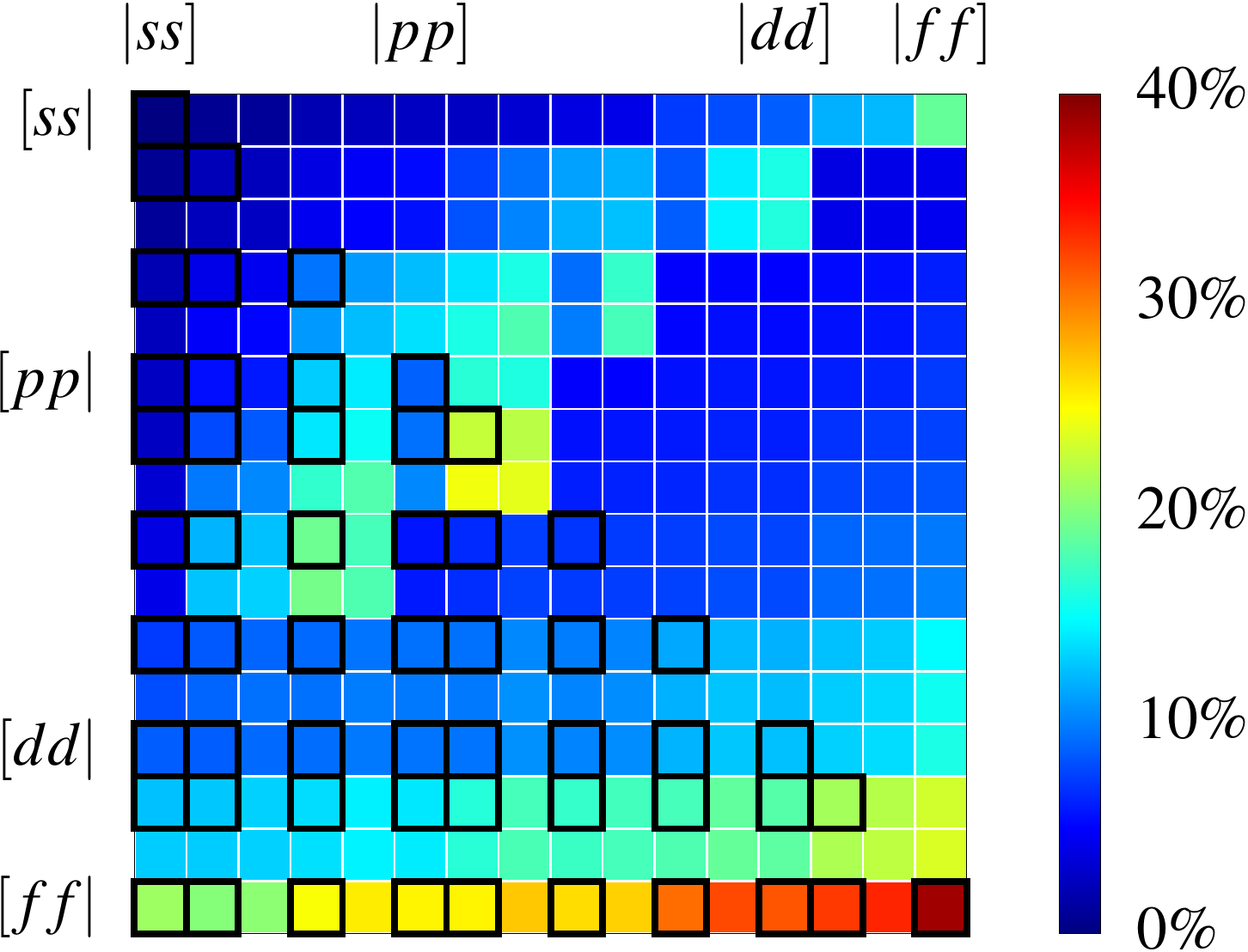}%
\label{fig:fpga_dsp}}
\hfil
\subfloat[$f_{\text{max}}$]{%
  \includegraphics[width=0.48\columnwidth]{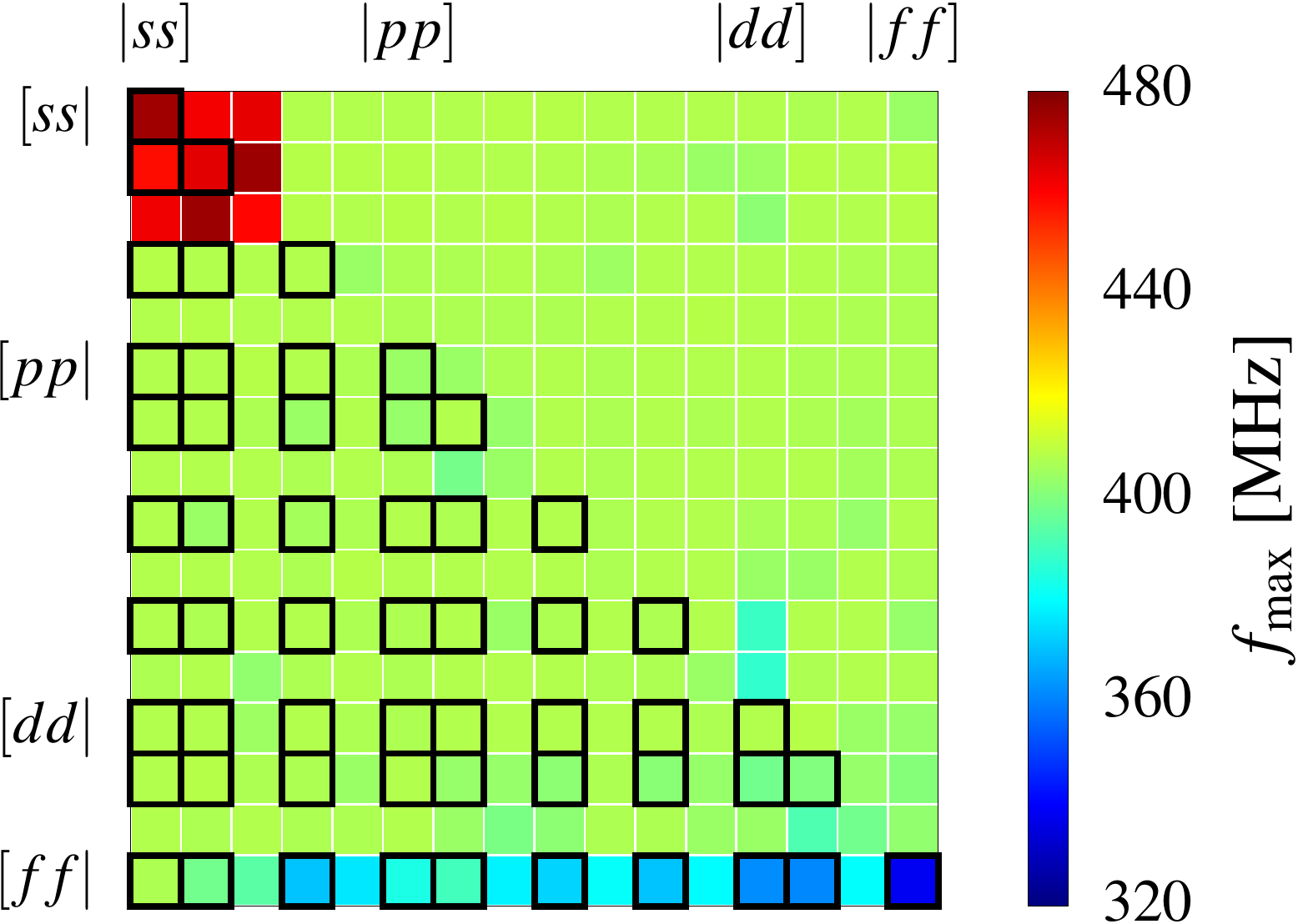}%
\label{fig:fpga_fmax}}
\caption{Resource utilization and clock frequency ($f_{\text{max}}$) of FPGA kernel designs for 256 \emph{generic} ERI quartet classes.
The 55 \emph{canonical} quartet classes are highlighted in black squares.
Available ALMs, registers, and DSPs in Stratix 10 GX 2800 are listed.}
\label{fig:fpga_resources}
\end{figure}

Fig.~\ref{fig:fpga_resources} shows FPGA resource utilization and clock frequency ($f_{\text{max}}$) for the 256 kernels with the optimal $c_{\text{max}}$.
Quartet classes $[ab|cd]$ are arranged as square element in heatmap matrices with row and column denoting $[ab|$ and $|cd]$, respectively.
Hence, $[ss|ss]$ forms the top left of heatmap matrices, $[ff|ss]$ the bottom left and $[ff|ff]$ the bottom right.
The 55 \emph{canonical} quartet classes, all located in the lower triangular of heatmap matrices, are highlighted.
The highest resource consumption is found for $[ff|**]$ designs, since the corresponding loops ($i$, $j$, $k$ in Fig.~\ref{fig:fpga_algorithm} line~\ref{lst:line:forall_rr}, $a$, $b$ in Fig.~\ref{fig:fpga_algorithm} line~\ref{lst:line:forall_quad}) are always unrolled, followed by designs below $[dd|**]$ (i.e.\ $[fd|**]$, $[df|**]$) that also have a high degree of parallelism in these loops. %
The usage of DSPs and $f_{\text{max}}$ are indicators for the number of arithmetic operations and design complexity, e.g.\ the simplest $[ss|ss]$ kernel only uses 8 DSPs and reaches $f_{\text{max}}=$ 474 MHz. %
In contrast, $[ff|ff]$ uses 2182 (39\%) DSPs for up to 3227 floating-point operations per cycle
at $f_{\text{max}}=$ 338 MHz. %

\begin{figure}[tbh]
\centering
\includegraphics[width=0.98\columnwidth]{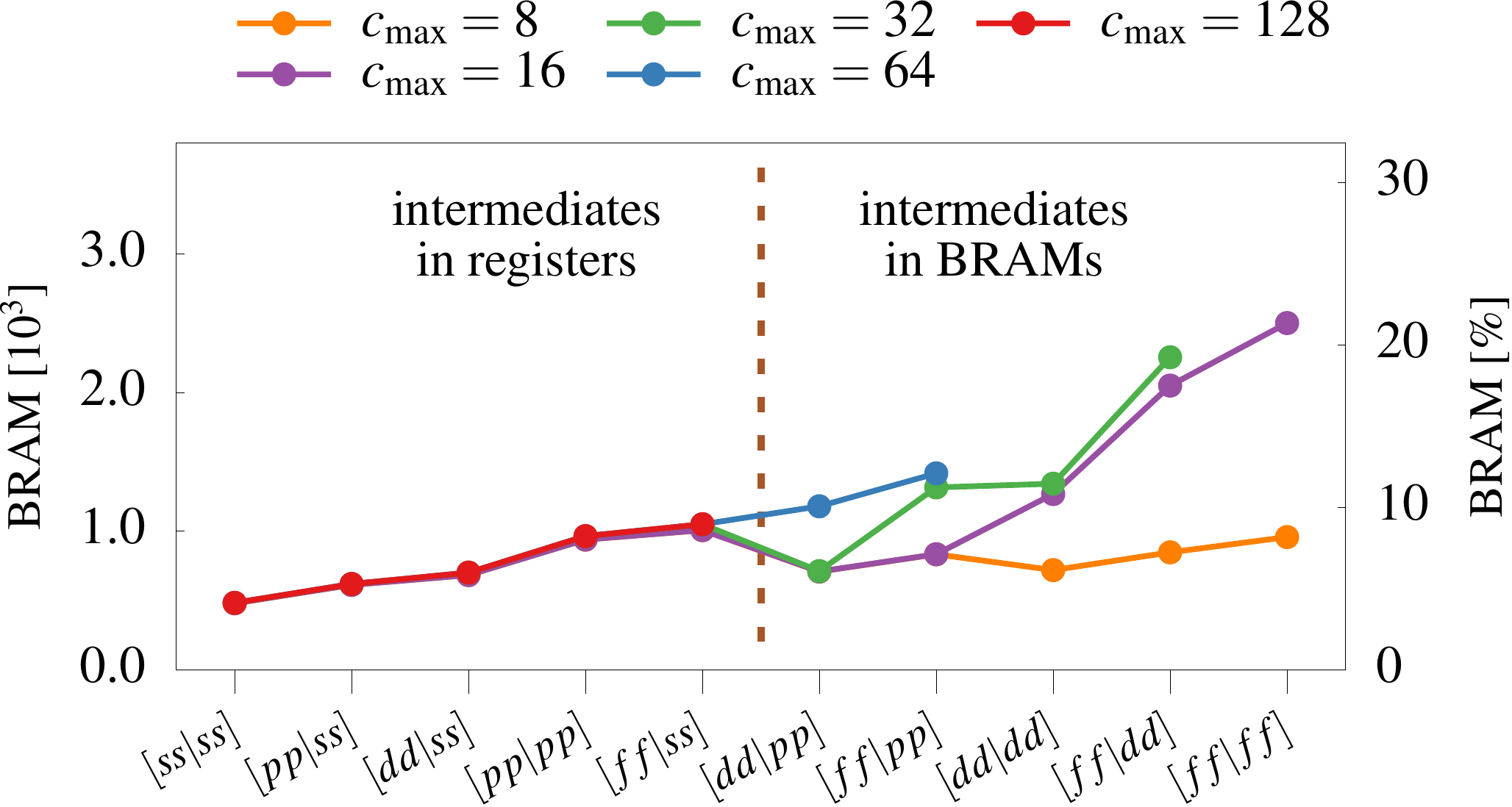}
\caption{BRAM utilization of FPGA kernels with doubled $c_{\text{max}}$ for selected \emph{canonical} $[aa|cc]$ quartet classes.}
\label{fig:fpga_bram}
\end{figure}

In Fig.~\ref{fig:fpga_bram}, BRAM utilization for selected \emph{canonical} quartet classes from $[ss|ss]$ to $[ff|ff]$ is plotted with respect to $c_{\text{max}}$ from 8 up to the optimal value. %
Depending on the depths of the local memory layout,
BRAM utilization may or may not increase with $c_{\text{max}}$ to create space for more private copies. %
It turns out that
the optimal $c_{\text{max}}$ is small for large quartet classes that already perform many iterations in inner loops. In contrast, for intermediate quartet classes with their shallow intermediate buffers, many private copies fit into the otherwise underutilized BRAMs.
Thus, the overall BRAM utilization is always less than 2.5K (21\%) in Stratix 10 GX 2800.

\begin{figure*}
\centering
\includegraphics[width=0.98\textwidth]{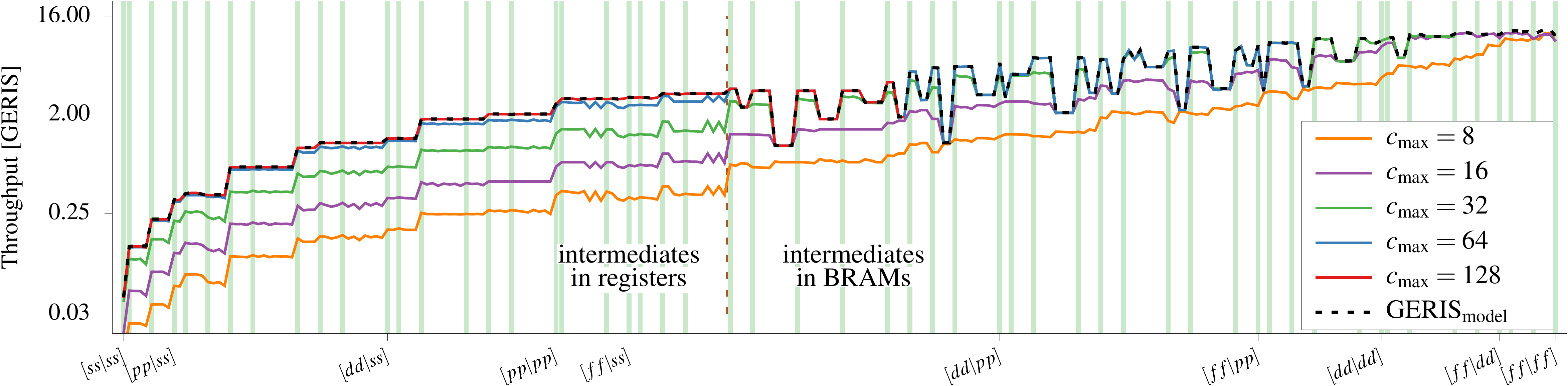}
\caption{Measured FPGA kernel throughput up to the optimal $c_{\text{max}}$ and comparison to the performance model for 256 \emph{generic} ERI quartet classes.
Throughput is plotted in $\log_{2}$-scale for visualizing doubled GERIS with respect to doubled $c_{\text{max}}$.
 The \emph{canonical} ERI quartet classes are highlight by vertical green bars.}
\label{fig:perf_model}
\end{figure*}

Fig.~\ref{fig:perf_model} presents the measured throughput of FPGA kernels for 256 \emph{generic} ERI quartet classes for different $c_{\text{max}}$ values up to the optimum.
For small to medium quartet classes, the throughput is only saturated with $c_{\text{max}}=128$. For the designs with all intermediates in registers, throughput exactly doubles with every doubling of $c_{\text{max}}$ before the last step.
In general, the throughput for large quartet classes, i.e.\ using BRAMs for intermediates, also increases when doubling $c_{\text{max}}$, but gets saturated more quickly with the best throughput reaching 10 - 11 GERIS. 

In order to compare the performance model motivated in Sec.~\ref{paragraph:default::initial_perf} with the measurements, we include the final $f_{\text{max}}$ that is automatically set by the synthesis tools after timing analysis. The modeled GERIS throughput is calculated as 
\begin{IEEEeqnarray}{lCr}
  \text{GERIS}_{\text{model}} = \frac{f_{\text{max}} \cdot n_{\text{ERIQ}}}%
  {\max(n_{\text{RR}}, n_{\text{GQ}}, n_{\text{CS}}) + 10},
\end{IEEEeqnarray}
with the denominator representing the number of clock cycles spent per quartet. Note that this model does not contain an explicit term for off-chip bandwidth, since this is implicitly encoded in the $n_{\text{CS}}$ term for the \emph{compress-store loops} that already incorporates the effect of output padding, as discussed in Sec.~\ref{sec:general_optimization}. However, there is an overhead per iteration of the outermost \emph{quartets loop} that we empirically determined as 10 clock, which presumably is related to the global memory operations on the interleaved memory interface.
This throughput model, as overlaid with the measurements in Fig.~\ref{fig:perf_model} is evidently in excellent agreement with the measurements for optimal $c_{\text{max}}$, validating the outcome of this optimization process.

\subsection{Preference for Canonical Classes}\label{subsec:canonical}

As introduced in Sec.~\ref{subsec:bg:quartets}, due to permutation symmetry, there can be up to eight \emph{generic} quartet classes that are mathematically identical and when reordering the inputs can be used for the same calculations. There are 55 unique \emph{canonical} classes that follow the convention defined in Eq.~\ref{eq:can_abcd}. With suitable pre-processing, these are sufficient to perform all ERI calculations discussed here.

With the loop structure presented in Fig.~\ref{fig:fpga_algorithm}, we ordered the loops in such a way that the innermost loops that are unrolled first correspond to the larger dimensions (Eq.~\ref{eq:can_abcd}) in the \emph{canonical} classes and thus contain more parallelism after the general optimization as presented in Sec.~\ref{sec:general_optimization}. With the automatic further unrolling as presented in Sec.~\ref{sec:further_unrolling} the designs for non-canonical classes can conceptually catch up to sufficient levels of parallelism, when the local memory layout permits.

In Fig.~\ref{fig:perf_model}, we highlighted the \emph{canonical} classes, each of which is followed by its \emph{generic} permutations to its right side. We see that for the designs with all intermediates in registers, for the optimal $c_{\text{max}}$ values all permutations match the performance of their respective \emph{canonical} representations. For the designs with intermediates in BRAMs, there are a number of permutations that do not reach the performance of their \emph{canonical} representations because of local memory limitations preventing further unrolling. 
Table~\ref{tab:permut_variants} provides another perspective on this effect for a few selected quartet classes, comparing the designs for three \emph{canonical} classes each with one of their \emph{generic} counterparts. For the first two examples, the respective \emph{canonical} versions reach the higher performance in accordance with their higher parallelism reflected by the performance model and despite the clock frequency advantage of the $[dp|ff]$ design with its lower resource utilization. 
\begin{table}
\renewcommand{\arraystretch}{1.3}
\caption{Comparison of some permutation variants for representative ERI quartet classes. $\max(n_{\text{RR}}, n_{\text{GQ}}, n_{\text{CS}})$ in Bold.}
\label{tab:permut_variants}
\centering
\begin{tabular}{@{\;}ccc@{\;\;\;}cccc@{\;\;\;}c@{\;}}
\hline
\multirow{2}{*}{Quartet}              &
$f_{\text{max}}$                      &
\multirow{2}{*}{$n_{\text{ERIQ}}$}    &
\multicolumn{4}{c}{Performance model} &
measured                              \\
\cmidrule(r){4-7}
& [MHz] & & $n_{\text{RR}}$ & $n_{\text{GQ}}$ & $n_{\text{CS}}$ & GERIS & GERIS \\
\hline
$[fd|ps]\,^{1}$ & 408.3 & \multirow{2}{*}{180}  & \textbf{12} &          3  & 6 & 3.34 & 3.33 \\
$[ps|fd]$       & 408.2 &                       &         12  & \textbf{60} & 6 & 1.05 & 1.05 \\
\hline
$[ff|dp]\,^{1}$ & 373.4 & \multirow{2}{*}{1800} & 30 &          18  & \textbf{57} & 10.03 & 10.03 \\
$[dp|ff]$       & 407.7 &                       & 60 & \textbf{100} &         57  &  6.67 &  6.65 \\
\hline
$[fd|fd]\,^{1}$ & 400.2 & \multirow{2}{*}{3600} & 72 &   60 & \textbf{113} & 11.71 & 11.17 \\
$[df|fd]$       & 392.5 &                       & 72 &   60 & \textbf{113} & 11.49 & 10.93 \\
\hline
\multicolumn{7}{l}{\footnotesize$^1$\,\emph{Canonical} ERI quartet classes.}
\end{tabular}
\end{table}

\subsection{Comparison with CPUs}\label{subsec:vs_libint}
We compare the performance of the FPGA design which includes compression to the performance of libint (version 2.7.2) without compression. Libint is a highly tuned and widely used CPU library for ERI computation using the algorithms based on the Obara-Saika method~\cite{Obara86, Obara88}.
Double-precision floating-point arithmetic is utilized for ERI computation in libint and these values are used as the numerical reference for our FPGA kernels for the purpose of the accuracy investigation.%
\begin{figure}
\centering
\includegraphics[width=0.98\columnwidth]{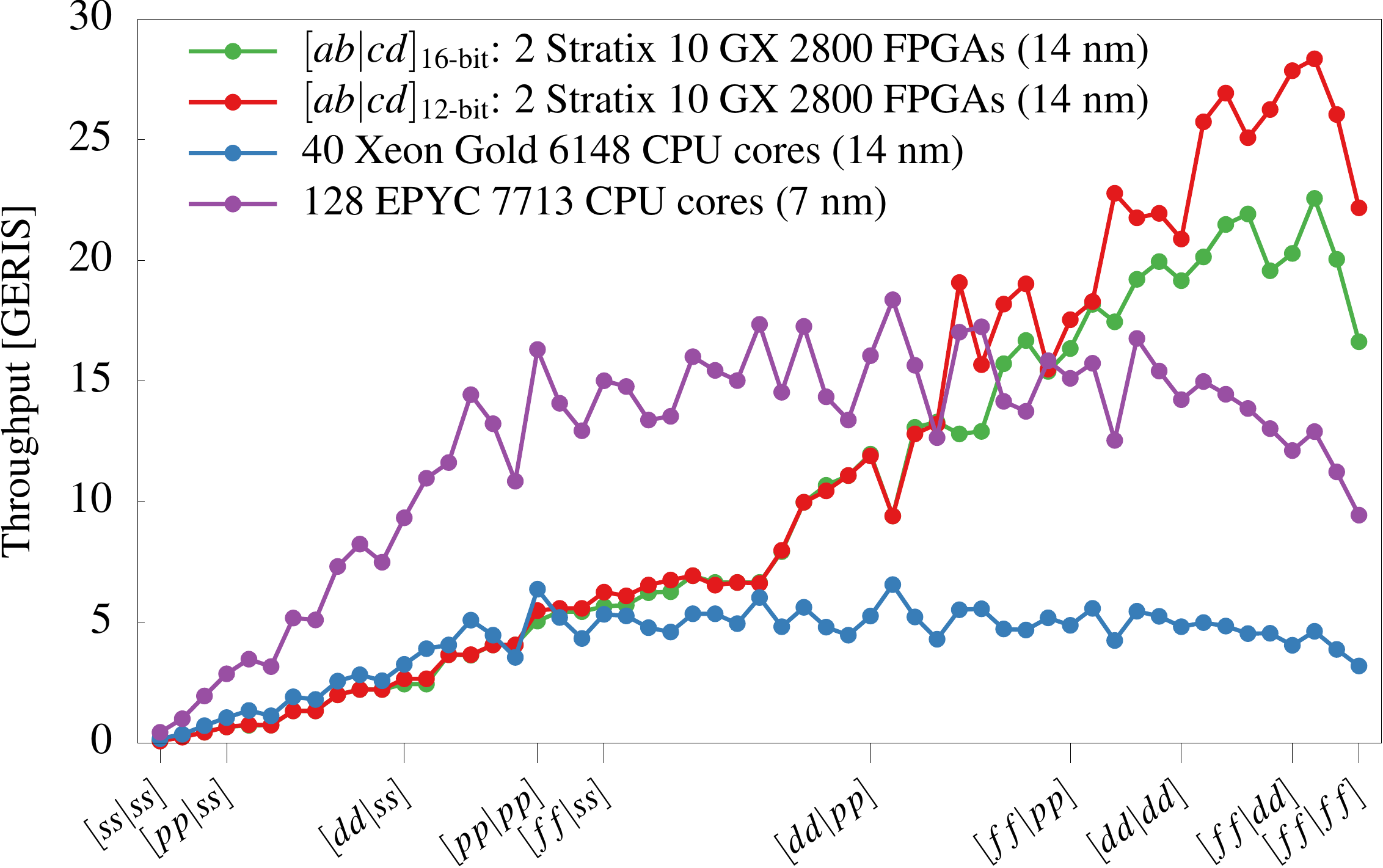}
\caption{Performance comparison between libint on CPUs (blue and purple lines) and FPGA kernels for \emph{canonical} $[ab|cd]_{16\text{-bit}}$ and $[ab|cd]_{12\text{-bit}}$.}
\label{fig:all_geris}
\end{figure}
A test program using the recommended modern C++ API of libint is used and parallelized with MPI, so that all ERIs of the same $[ab|cd]$ quartet class in the benchmark molecule are distributed to different ranks and computed in parallel. GCC (version 11.2.0) and OpenMPI (version 4.1.1) are used to build libint and the test program.
We carried out the measurements on a two-socket server with Intel Xeon Gold 6148 CPUs (2x20 cores) as well as a two-socket server with AMD EPYC 7713 CPUs (2x64 cores). The Intel Xeon CPU represents a CPU reference using the same manufacturing technology as the FPGAs, with both devices showing up first in 2018 in HPC centers, and the AMD EPYC CPU is chosen as a representative for up-to-date manufacturing technology.
For Xeon Gold 6148 CPU and EPYC 7713 CPU the compiler flags \texttt{-march=skylake-avx512} and \texttt{-march=znver3} are used, respectively, for optimized performance. For the FPGA kernel execution, two Stratix 10 GX 2800 FPGA cards in one compute node are used with an OpenMP-parallelized driver code that uses two SYCL queues, one per FPGA.

Fig.~\ref{fig:all_geris} presents the results of measured throughput in GERIS for the 55 \emph{canonical} ERI quartet classes on CPUs and FPGA cards.
The throughput for $[ss|ss]$ is merely 0.08~GERIS for both 12-bit and 16-bit FPGA kernels, whereas it is 0.17~GERIS on 40 Intel Xeon CPU cores and 0.48~GERIS on 128 AMD EPYC CPU cores. For higher angular momenta, the number of ERIs per quartet, $n_{\text{ERIQ}}$, as well as the number of floating-point operations per quartet, $n_{\text{FLOPQ}}$, increases. As a consequence, the throughput on both CPUs and FPGAs starts to increase.
Before $[ff|ss]$ ($n_{\text{ERIQ}} = 100$) 40 Intel Xeon CPU cores slightly outperform the FPGA kernels.
Thereafter, the throughput of FPGA kernels increases steeply and can outperform the 40 Intel Xeon CPU cores by factors up to 6x. This shows the advantage of the FPGA design over a CPU 
using the same 14 nm manufacturing technology.
When compared with 128 AMD EPYC CPU cores, released in 2021 and produced with a much more recent 7 nm manufactoring technology, the FPGA kernels start take over the performance lead later, around $[ff|ds]$ ($n_{\text{ERIQ}} = 600$) with \ 14.1~GERIS on CPUs, and 15.3~GERIS and 18.2~GERIS for 16-bit and 12-bit FPGA kernels, respectively.
For even larger ERI quartet classes, the throughput in terms of GERIS on 128 AMD EPYC CPU cores starts to decrease. %
In contrast, the throughput of both 16-bit and 12-bit FPGA kernels continuously increases and reaches the peaks of 22.3~GERIS and 28.4~GERIS for 16-bit and 12-bit kernels, respectively.
When comparing the throughput of 16-bit and 12-bit FPGA kernels, similar performance is found for ERI quartet classes smaller than $[fd|pp]$ ($n_{\text{ERIQ}} = 540$), then the 12-bit FPGA kernel starts to outperform the 16-bit kernels in general.

\begin{figure}
\centering
\includegraphics[width=0.98\columnwidth]{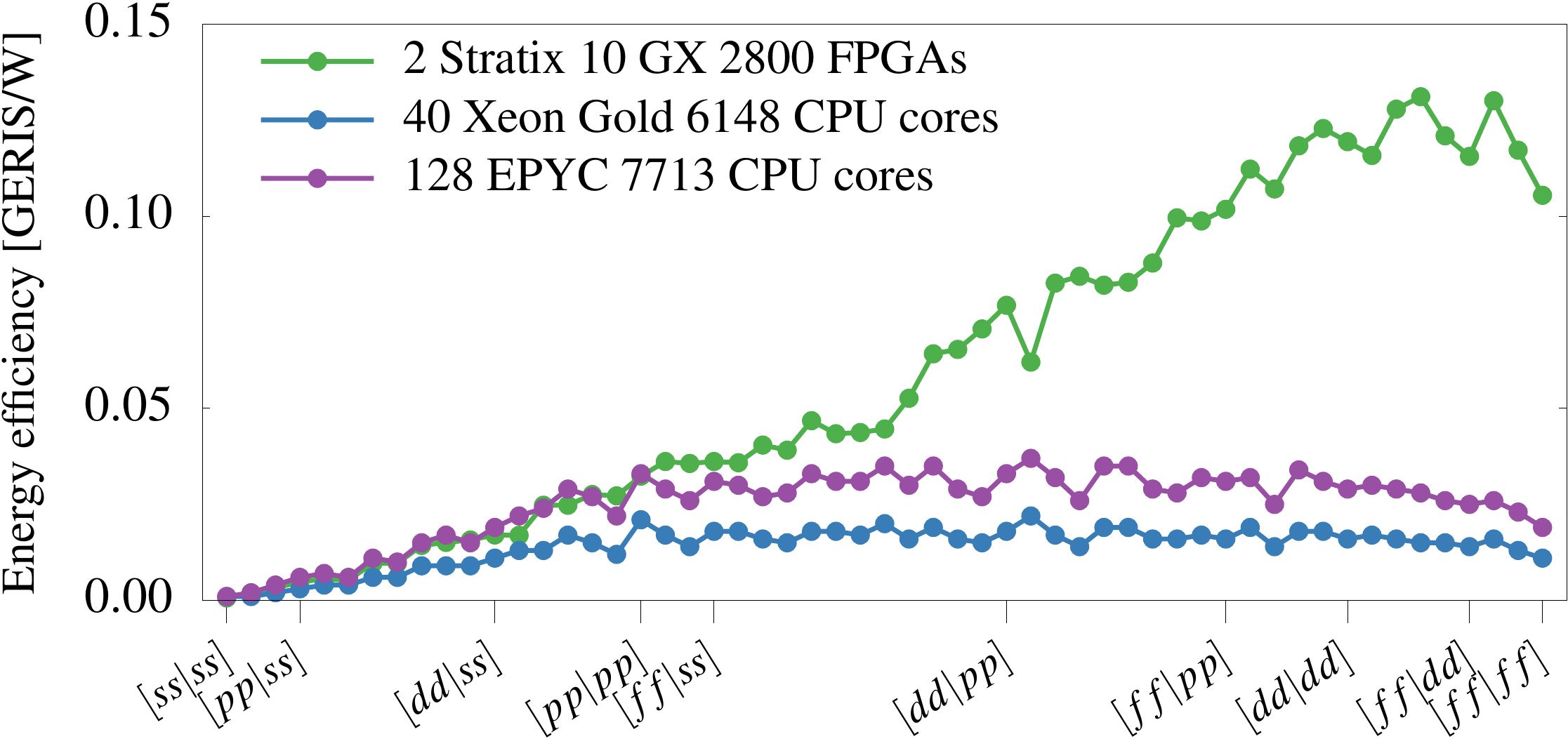}
\caption{Comparison of energy efficiency between libint on CPUs (blue and purple lines) and FPGA kernels for \emph{canonical} $[ab|cd]_{16\text{-bit}}$.}
\label{fig:energy_eff}
\end{figure}
During the benchmark runs, the CPU package power consumption measured with RAPL counters was nearly independent of the type of quartet. The power consumption was about 490~W for two AMD EPYC 7713 CPUs (TDP 240 W each) and 300~W for two Intel Xeon Gold 6148 CPUs (TDP 150 W each). In contrast, for two FPGA cards, the combined board power consumption starts of from 132~W for $[ss|ss]$ and increases to 180~W for $[ff|ff]$. The resulting energy efficiency is shown in Fig.~\ref{fig:energy_eff}. For quartets with low angular momenta, the power efficiency of CPUs and FPGAs is comparable, while for intermediate to larger angular momenta there is a clear advantage of the newer AMD Zen 3 CPU cores over the older Intel Skylake-SP CPU cores. Starting from $[pp|pp]$ the advantage of the FPGA implementation over the CPUs becomes evident, reaching about 5x higher energy efficiency of the FPGA implementation for high angular momenta.

\begin{figure}
\centering
\includegraphics[width=0.98\columnwidth]{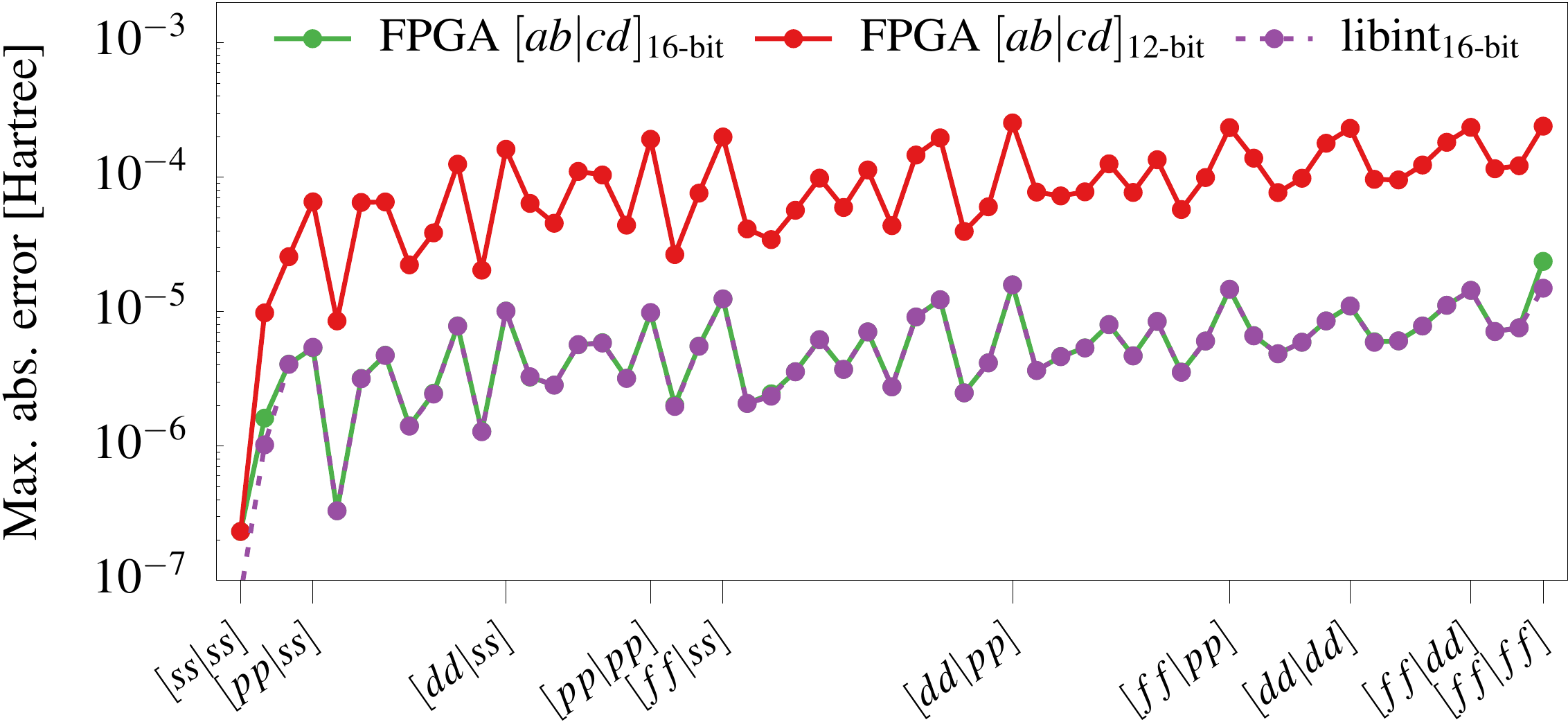}
\caption{Comparison of maximum absolute error for decompressed ERIs using 16-bit integer for libint, 16- and 12-bit integers of FPGA kernels.}
\label{fig:num_acc}
\end{figure}
Last but not least, the numerical accuracy due to ERI compression is also considered and compared against the libint reference in double-precision.
We implemented an ERI compression using 16-bit integer based on the libint reference values.
Fig.~\ref{fig:num_acc} shows the maximum absolute error for decompressed ERIs for the \emph{canonical} quartet classes.
If 16-bit integer is adopted for ERI compression, libint and FPGA kernel give almost identical max.\ abs.\ errors, $10^{-7}$ - $10^{-5}$ Hartree, because the errors are bound by $\epsilon/2$ (Eq.~\ref{eq:decomp_eris}), irrespective of single-precision or double-precision floating-point arithmetics for ERI computation.
With 12-bit integer for ERI compression, max.\ abs.\ errors increase systematically and, except for $[ss|ss]$, range from $10^{-5}$ - $10^{-4}$ Hartree.
Hence, the main source of numerical errors is attributed to the bitwidth used for ERI compression.

\section{Related Work}\label{sec:related_work}

In 2008, Kindratenko et al.~\cite{ki-uf-a} presented a first attempt for ERI calculation on FPGAs using the SRC MAP C compiler. The evaluation was limited to $[ss|ss]$ quartets, which forms the simplest, and due to its limited computational complexity also little promising point in the design space covered in this work. The main contribution of Kindratenko et al.~\cite{ki-uf-a} was to improve pipelining through the manual fusion of selected nested loops. A similar approach was presented for other kinds of two electron integrals used for studying electron scattering~\cite{gi-st-10a,gi-st-12a} using Mitrion-C. Loop fusion shows up in our work as attribute guided automatic loop coalescing in Sec.~\ref{paragraph:default:rrloops} and Sec.~\ref{paragraph:default:gqloops} after customization of parallelism.

\section{Conclusion}\label{sec:conclusion}

In this work, we implement the first ERI computation and compression with arbitrary bitwidth integer on FPGAs. %
Targeting 256 ERI quartet classes up to $L=3$ with their diverse computation characteristics, high throughput is achieved via FPGA kernels that are customized at compile-time via function templates by using the Intel oneAPI DPC++ compiler.
The measured throughput is analyzed and accurately explained by a performance model.
Our evaluation shows that the FPGA kernels parallelized on 2 Stratix 10 GX 2800 outperform libint, an optimized library for ERI computation, by factors up to 6.0x on two-socket server with 40 Intel Xeon Gold 6148 CPU cores and up to 1.9x on 128 AMD EPYC 7713 CPU cores.
Meanwhile, up to 5x energy efficiency is reached for the benchmark computation compared to the newer CPUs.
The numerical errors are mainly due to the bitwidth used in ERI compression, regardless of single- or double-precision arithmetic operations used in ERI computation.

The ERI computation is a key component in hybrid-DFT based AIMD simulations. This work paves the way for FPGA-accelerated quantum-mechanics-based atomic simulations.
Further development in the direction of a suitable libint-like library interface allowing to use of the implementation in many quantum chemistry codes is already on the way.

\section*{Acknowledgment}
The authors gratefully acknowledge the computing time provided to them on the high-performance computer Noctua~2 at the NHR Center PC$^2$. These are funded by the Federal Ministry of Education and Research and the state governments participating on the basis of the resolutions of the GWK for the national high performance computing at universities (www.nhr-verein.de/unsere-partner).

\bibliographystyle{IEEEtran}
\bibliography{bare_conf.bib}

\begin{thebibliography}{10}
\providecommand{\url}[1]{#1}
\csname url@samestyle\endcsname
\providecommand{\newblock}{\relax}
\providecommand{\bibinfo}[2]{#2}
\providecommand{\BIBentrySTDinterwordspacing}{\spaceskip=0pt\relax}
\providecommand{\BIBentryALTinterwordstretchfactor}{4}
\providecommand{\BIBentryALTinterwordspacing}{\spaceskip=\fontdimen2\font plus
\BIBentryALTinterwordstretchfactor\fontdimen3\font minus
  \fontdimen4\font\relax}
\providecommand{\BIBforeignlanguage}[2]{{%
\expandafter\ifx\csname l@#1\endcsname\relax
\typeout{** WARNING: IEEEtran.bst: No hyphenation pattern has been}%
\typeout{** loaded for the language `#1'. Using the pattern for}%
\typeout{** the default language instead.}%
\else
\language=\csname l@#1\endcsname
\fi
#2}}
\providecommand{\BIBdecl}{\relax}
\BIBdecl

\bibitem{Kuehne2014}
T.~D. K{\"u}hne, ``Second generation {Car-Parrinello} molecular dynamics,''
  \emph{WIREs Comput. Mol. Sci.}, vol.~4, no.~4, pp. 391--406, 2014.

\bibitem{pople98}
J.~A. Pople, ``Nobel lecture: Quantum chemical models,'' \emph{Rev. Mod.
  Phys.}, vol.~71, no.~5, pp. 1267--1274, 1999.

\bibitem{kohn98}
W.~Kohn, ``Nobel lecture: Electronic structure of matter---wave functions and
  density functionals,'' \emph{Rev. Mod. Phys.}, vol.~71, no.~5, pp.
  1253--1266, 1999.

\bibitem{Lass20}
M.~Lass, R.~Schade, T.~D. Kühne, and C.~Plessl, ``A submatrix-based method for
  approximate matrix function evaluation in the quantum chemistry code
  {CP2K},'' in \emph{SC20: International Conference for High Performance
  Computing, Networking, Storage and Analysis}, 2020, pp. 1--14.

\bibitem{Schade22}
R.~Schade, T.~Kenter, H.~Elgabarty, M.~Lass, O.~Schütt, A.~Lazzaro, H.~Pabst,
  S.~Mohr, J.~Hutter, T.~D. Kühne, and C.~Plessl, ``Towards electronic
  structure-based ab-initio molecular dynamics simulations with hundreds of
  millions of atoms,'' \emph{Parallel Computing}, vol. 111, p. 102920, 2022.

\bibitem{Becke93}
A.~D. Becke, ``A new mixing of {Hartree-Fock} and local density-functional
  theories,'' \emph{The Journal of Chemical Physics}, vol.~98, no.~2, pp.
  1372--1377, 1993.

\bibitem{PhysRevApplied.8.024023}
A.~Sharan, Z.~Gui, and A.~Janotti, ``Hybrid-functional calculations of the
  copper impurity in silicon,'' \emph{Phys. Rev. Appl.}, vol.~8, p. 024023, Aug
  2017.

\bibitem{Kuehne2020}
T.~D. K{\"u}hne, M.~Iannuzzi, M.~Del~Ben, V.~V. Rybkin, P.~Seewald, F.~Stein,
  T.~Laino, R.~Z. Khaliullin, O.~Sch{\"u}tt, F.~Schiffmann \emph{et~al.},
  ``Cp2k: An electronic structure and molecular dynamics software
  package-quickstep: Efficient and accurate electronic structure
  calculations,'' \emph{J. Chem. Phys.}, vol. 152, no.~19, p. 194103, 2020.

\bibitem{HJO_ch9}
T.~Helgaker, P.~Jørgensen, and J.~Olsen, \emph{Molecular Integral
  Evaluation}.\hskip 1em plus 0.5em minus 0.4em\relax John Wiley \& Sons, Ltd,
  2000, ch.~9, pp. 336--432.

\bibitem{libint2}
E.~F. Valeev, ``Libint: A library for the evaluation of molecular integrals of
  many-body operators over {Gaussian} functions,'' http://libint.valeyev.net,
  2022, version 2.7.2.

\bibitem{libcint}
Q.~Sun, ``Libcint: An efficient general integral library for {Gaussian} basis
  functions,'' \emph{Journal of Computational Chemistry}, vol.~36, pp.
  1664--1671, 2015.

\bibitem{Chow16}
B.~P. Pritchard and E.~Chow, ``Horizontal vectorization of electron repulsion
  integrals,'' \emph{Journal of Computational Chemistry}, vol.~37, no.~28, pp.
  2537--2546, 2016.

\bibitem{Chow18}
H.~Huang and E.~Chow, ``Accelerating quantum chemistry with vectorized and
  batched integrals,'' in \emph{SC18: International Conference for High
  Performance Computing, Networking, Storage and Analysis}, 2018, pp. 529--542.

\bibitem{Neese22}
F.~Neese, ``The {SHARK} integral generation and digestion system,''
  \emph{Journal of Computational Chemistry}, 2022.

\bibitem{Martínez08}
I.~S. Ufimtsev and T.~J. Martínez, ``Quantum chemistry on graphical processing
  units. 1. strategies for two-electron integral evaluation,'' \emph{Journal of
  Chemical Theory and Computation}, vol.~4, no.~2, pp. 222--231, 2008.

\bibitem{Gordon10}
A.~Asadchev, V.~Allada, J.~Felder, B.~M. Bode, M.~S. Gordon, and T.~L. Windus,
  ``Uncontracted {Rys} quadrature implementation of up to {G} functions on
  graphical processing units,'' \emph{Journal of Chemical Theory and
  Computation}, vol.~6, no.~3, pp. 696--704, 2010.

\bibitem{Merz13}
Y.~Miao and K.~M.~J. Merz, ``Acceleration of electron repulsion integral
  evaluation on graphics processing units via use of recurrence relations,''
  \emph{Journal of Chemical Theory and Computation}, vol.~9, no.~2, pp.
  965--976, 2013.

\bibitem{Gordon22}
J.~L.~G. Vallejo, G.~M. Barca, and M.~S. Gordon, ``High-performance
  {GPU}-accelerated evaluation of electron repulsion integrals,''
  \emph{Molecular Physics}, p. e2112987, 2022.

\bibitem{DeMatteis20}
T.~De~Matteis, J.~de~Fine~Licht, and T.~Hoefler, ``{FBLAS}: Streaming linear
  algebra on {FPGA},'' in \emph{Proceedings of the International Conference for
  High Performance Computing, Networking, Storage and Analysis}, ser. SC
  '20.\hskip 1em plus 0.5em minus 0.4em\relax IEEE Press, 2020.

\bibitem{Gorlani19}
P.~Gorlani, T.~Kenter, and C.~Plessl, ``{OpenCL} implementation of {Cannon's}
  matrix multiplication algorithm on {Intel} {Stratix} 10 {FPGAs},'' in
  \emph{2019 International Conference on Field-Programmable Technology
  (ICFPT)}, 2019, pp. 99--107.

\bibitem{Meyer22}
M.~Meyer, T.~Kenter, and C.~Plessl, ``In-depth {FPGA} accelerator performance
  evaluation with single node benchmarks from the {HPC} challenge benchmark
  suite for {Intel} and {Xilinx} {FPGAs} using {OpenCL},'' \emph{Journal of
  Parallel and Distributed Computing}, vol. 160, pp. 79--89, 2022.

\bibitem{Zhang22}
Y.~Du, Y.~Hu, Z.~Zhou, and Z.~Zhang, ``High-performance sparse linear algebra
  on {HBM}-equipped {FPGAs} using {HLS}: A case study on {SpMV},'' in
  \emph{Proceedings of the 2022 ACM/SIGDA International Symposium on
  Field-Programmable Gate Arrays}, ser. FPGA '22.\hskip 1em plus 0.5em minus
  0.4em\relax New York, NY, USA: Association for Computing Machinery, 2022, pp.
  54--64.

\bibitem{Herbordt05}
Y.~Gu, T.~VanCourt, and M.~C. Herbordt, ``Accelerating molecular dynamics
  simulations with configurable circuits,'' in \emph{International Conference
  on Field Programmable Logic and Applications, 2005.}, 2005, pp. 475--480.

\bibitem{Herbordt10}
M.~Chiu and M.~C. Herbordt, ``Molecular dynamics simulations on
  high-performance reconfigurable computing systems,'' \emph{ACM Trans.
  Reconfigurable Technol. Syst.}, vol.~3, no.~4, nov 2010.

\bibitem{Herbordt13}
M.~A. Khan, M.~Chiu, and M.~C. Herbordt, \emph{{FPGA}-Accelerated Molecular
  Dynamics}.\hskip 1em plus 0.5em minus 0.4em\relax New York, NY: Springer New
  York, 2013, pp. 105--135.

\bibitem{Herbordt19}
C.~Yang, T.~Geng, T.~Wang, R.~Patel, Q.~Xiong, A.~Sanaullah, C.~Wu, J.~Sheng,
  C.~Lin, V.~Sachdeva, W.~Sherman, and M.~Herbordt, ``Fully integrated {FPGA}
  molecular dynamics simulations,'' in \emph{Proceedings of the International
  Conference for High Performance Computing, Networking, Storage and Analysis},
  ser. SC '19.\hskip 1em plus 0.5em minus 0.4em\relax New York, NY, USA:
  Association for Computing Machinery, 2019.

\bibitem{Jones22}
D.~Jones, J.~E. Allen, Y.~Yang, W.~F. Drew~Bennett, M.~Gokhale, N.~Moshiri, and
  T.~S. Rosing, ``Accelerators for classical molecular dynamics simulations of
  biomolecules,'' \emph{Journal of Chemical Theory and Computation}, vol.~18,
  no.~7, pp. 4047--4069, 2022.

\bibitem{ki-uf-a}
V.~Kindratenko, I.~Ufimtsev, and T.~Martínez, ``Evaluation of two-electron
  repulsion integrals over gaussian basis functions on src-6 reconfigurable
  computer,''
  \url{https://users.ncsa.illinois.edu/kindr/papers/rssi08\_paper2.pdf}, 2008.

\bibitem{FPGAOpt4oneAPI}
\BIBentryALTinterwordspacing
\emph{{FPGA} Optimization Guide for Intel oneAPI Toolkits: Developer Guide},
  Intel, September 2022. [Online]. Available:
  \url{https://www.intel.com/content/www/us/en/develop/documentation}
\BIBentrySTDinterwordspacing

\bibitem{libintwiki}
``libint wiki,''
  \url{https://github.com/evaleev/libint/wiki\#program-specific-notes},
  accessed: 2022-12-15.

\bibitem{zenodo}
\BIBentryALTinterwordspacing
X.~Wu, T.~Kenter, R.~Schade, T.~D. Kühne, and C.~Plessl, ``{Source Code for
  Computing and Compressing Electron Repulsion Integrals on FPGAs},'' Mar.
  2023. [Online]. Available: \url{https://doi.org/10.5281/zenodo.7763841}
\BIBentrySTDinterwordspacing

\bibitem{Kenny08}
J.~P. Kenny, C.~L. Janssen, E.~F. Valeev, and T.~L. Windus, ``Components for
  integral evaluation in quantum chemistry,'' \emph{Journal of Computational
  Chemistry}, vol.~29, no.~4, pp. 562--577, 2008.

\bibitem{Boys50}
S.~F. Boys, ``Electronic wave functions - {I}. {A} general method of
  calculation for the stationary states of any molecular system,''
  \emph{Proceedings of the Royal Society of London. Series A. Mathematical and
  Physical Sciences}, vol. 200, no. 1063, pp. 542--554, 1950.

\bibitem{dupuis1976DKR}
M.~Dupuis, J.~Rys, and H.~F. King, ``Evaluation of molecular integrals over
  {Gaussian} basis functions,'' \emph{The Journal of Chemical Physics},
  vol.~65, no.~1, pp. 111--116, 1976.

\bibitem{king1976DKR}
H.~F. King and M.~Dupuis, ``Numerical integration using rys polynomials,''
  \emph{Journal of Computational Physics}, vol.~21, no.~2, pp. 144--165, 1976.

\bibitem{Rys83}
J.~Rys, M.~Dupuis, and H.~F. King, ``Computation of electron repulsion
  integrals using the {Rys} quadrature method,'' \emph{Journal of Computational
  Chemistry}, vol.~4, no.~2, pp. 154--157, 1983.

\bibitem{Obara86}
S.~Obara and A.~Saika, ``Efficient recursive computation of molecular integrals
  over {Cartesian} {Gaussian} functions,'' \emph{The Journal of Chemical
  Physics}, vol.~84, no.~7, pp. 3963--3974, 1986.

\bibitem{headgordon1988HGPRR}
M.~Head‐Gordon and J.~A. Pople, ``A method for two‐electron {Gaussian}
  integral and integral derivative evaluation using recurrence relations,''
  \emph{The Journal of Chemical Physics}, vol.~89, no.~9, pp. 5777--5786, 1988.

\bibitem{mcmurchie1978McMurchieDavidson}
L.~E. McMurchie and E.~R. Davidson, ``One- and two-electron integrals over
  {Cartesian} {Gaussian} functions,'' \emph{Journal of Computational Physics},
  vol.~26, no.~2, pp. 218--231, 1978.

\bibitem{Fülscher93}
M.~P. Fülscher and P.-O. Widmark, ``An electron repulsion integral compression
  algorithm,'' \emph{Journal of Computational Chemistry}, vol.~14, no.~1, pp.
  8--12, 1993.

\bibitem{Guidon08}
M.~Guidon, F.~Schiffmann, J.~Hutter, and J.~VandeVondele, ``Ab initio molecular
  dynamics using hybrid density functionals,'' \emph{The Journal of Chemical
  Physics}, vol. 128, no.~21, p. 214104, 2008.

\bibitem{Ying15}
J.~Lu and L.~Ying, ``Compression of the electron repulsion integral tensor in
  tensor hypercontraction format with cubic scaling cost,'' \emph{Journal of
  Computational Physics}, vol. 302, pp. 329--335, 2015.

\bibitem{Obara88}
S.~Obara and A.~Saika, ``General recurrence formulas for molecular integrals
  over {Cartesian} {Gaussian} functions,'' \emph{The Journal of Chemical
  Physics}, vol.~89, no.~3, pp. 1540--1559, 1988.

\bibitem{gi-st-10a}
C.~Gillan, T.~Steinke, J.~Bock, S.~Borchert, I.~Spence, and N.~Scott,
  ``Programming challenges for the implementation of numerical quadrature in
  atomic physics on fpga and gpu accelerators,'' in \emph{IEEE/ACM
  International Conference on Cluster, Cloud and Grid Computing}, 2010, pp.
  757--762.

\bibitem{gi-st-12a}
\BIBentryALTinterwordspacing
C.~J. Gillan, T.~Steinke, J.~Bock, S.~Borchert, I.~Spence, and N.~S. Scott,
  ``Comparing the implementation of two-dimensional numerical quadrature on
  gpu, fpga and clearspeed systems to study electron scattering by atoms,''
  \emph{Concurrency and Computation: Practice and Experience}, vol.~24, no.~1,
  pp. 84--95, 2012. [Online]. Available:
  \url{https://onlinelibrary.wiley.com/doi/abs/10.1002/cpe.1733}
\BIBentrySTDinterwordspacing

\end{thebibliography}

\end{document}